\documentclass[superscriptaddress,footinbib]{revtex4}

 \usepackage{graphicx}
 \usepackage{amssymb}
 \usepackage{amsthm}
 \usepackage{amsmath}
 \usepackage{mathtools}
 \usepackage{natbib}
 \usepackage{subeqnarray}
 \usepackage{color}
 \usepackage[english]{babel}
\usepackage{ifpdf}
\usepackage{psfrag}

\newcommand{\zip}[1]{ }

\def\@fnsymbol#1{\ifcase#1\or *\or \dagger\or \ddagger\or \mathchar "278\or \mathchar "27B\or \|\or **\or \dagger\dagger \or \ddagger\ddagger \else\@ctrerr\fi\relax}
\long\def\symbolfootnote[#1]#2{\begingroup%
\def\thefootnote{\fnsymbol{footnote}}\footnote[#1]{#2}\endgroup}

\begin{document}

\title{On the efficiency of energy harvesting using vortex-induced vibrations of cables}
%\author[LDX]{Kiran Singh\tnoteref{email}, S{\'e}bastien Michelin and Emmanuel de Langre}
\author{Cl{\'e}ment Grouthier}
\email{clement.grouthier@ladhyx.polytechnique.fr}
\affiliation{D{\'e}partement de M{\'e}canique, LadHyX,\\ CNRS -- {\'E}cole Polytechnique, 91128, Palaiseau, France}
\author{S{\'e}bastien Michelin}
\email{sebastien.michelin@ladhyx.polytechnique.fr}
\affiliation{D{\'e}partement de M{\'e}canique, LadHyX,\\ CNRS -- {\'E}cole Polytechnique, 91128, Palaiseau, France}
\author{R{\'e}mi Bourguet}
\email{bourguet@imft.fr}
\affiliation{Institut de M{\'e}canique des Fluides de Toulouse, Universit{\'e} de Toulouse and CNRS, 31400, Toulouse, France}
\author{Yahya Modarres-Sadeghi}
\email{modarres@engin.umass.edu}
\affiliation{Department of Mechanical and Industrial Engineering, University of Massachusetts, Amherst, MA 01003, USA}
\author{Emmanuel de Langre}
\email{delangre@ladhyx.polytechnique.fr}
\affiliation{D{\'e}partement de M{\'e}canique, LadHyX,\\ CNRS -- {\'E}cole Polytechnique, 91128, Palaiseau, France}
\date{\today}
\begin{abstract}

\noindent{Many technologies based on fluid-structure interaction mechanisms are being developed to harvest energy from geophysical flows. The velocity of such flows is low, and so is their energy density. Large systems are therefore required to extract a significant amount of energy. The question of the efficiency of energy harvesting using vortex-induced vibrations (VIV) of cables is addressed in this paper, through two reference configurations: (i) a long tensioned cable with periodically-distributed harvesters and (ii) a hanging cable with a single harvester at its upper extremity. After validation against either direct numerical simulations or experiments, an appropriate reduced-order wake-oscillator model is used to perform parametric studies of the impact of the harvesting parameters on the efficiency. For both configurations, an optimal set of parameters is identified and it is shown that the maximum efficiency is close to the value reached with an elastically-mounted rigid cylinder. The variability of the efficiency is studied in light of the fundamental properties of each configuration, i.e. body flexibility and gravity-induced spatial variation of the tension. In the periodically-distributed harvester configuration, it is found that the standing-wave nature of the vibration and structural mode selection play a central role in energy extraction. In contrast, the efficiency of the hanging cable is essentially driven by the occurrence of traveling wave vibrations.}

\end{abstract}

\maketitle

\section{Introduction}

Innovative energy harvesting devices are being developed to extract energy from geophysical flows such as wind or marine currents. The most common way to extract this energy is to convert it into the rotary motion of a dedicated structure, as wind turbines or marine turbines \citep{Nish,Baha,Batt}. Even if these technologies are now mature from an industrial point of view, several other mechanisms to harvest this energy are currently being studied. 

Among them, a specific class is based on flow-induced vibrations \citep{Blev,Naud}. Energy harvesting through fluid-elastic instabilities has been investigated, such as galloping \citep{Barr}, airfoil coupled-mode flutter \citep{Peng,Bora} and flutter in an axial flow \citep{Tang,Sing,Sing2,Mich}. Vortex-induced vibrations (VIV), a strong coupling between the solid dynamics and its fluctuating wake, are another interesting mechanism to extract energy from geophysical flows \citep{Yosh,Barr2}. \citet{Bern} for instance developed the VIVACE device to harvest energy by VIV of elastically-supported rigid cylinders.

A quantitative criterion is necessary to compare the performances of all these emerging technologies. As in \citet{Barr}, \citet{Bern} or \citet{Hobb}, we define the efficiency of the harvesting, $\eta$, as the ratio between the time-averaged extracted power $\left< \mathcal{P} \right>$, where $\left< . \right>$ stands for time-averaged quantities, and the energy flux across the cross-flow section $\mathcal{A}$ of the device, $\mathcal{P}_{0}$,

\begin{equation}
	\eta = \dfrac{\left< \mathcal{P} \right>}{\mathcal{P}_{0}} = \dfrac{\left< \mathcal{P} \right>}{\frac{1}{2} \rho \mathcal{A} U^{3}},
	\label{eq:efficiency}
\end{equation}
 
\noindent{where $\rho$ and $U$ are respectively the fluid density and the flow velocity. There are of course many other ways to define the efficiency of an energy harvesting device, see for instance \citet{Zhu} or \citet{Doar}. In such geophysical flows, the energy density ($ \rho U^{3} /2$) is low, of the order of 500 $\mbox{W/m}^{2}$ for typical wind speed (10 $\mbox{m/s}$) or current speed (1 $\mbox{m/s}$). A large area $\mathcal{A}$ is thus needed to access large quantitites of energy. This can be achieved either by considering many short devices, as in \citet{Bern}, or one single large structure, as considered in the present paper. 

In the specific domain of VIV, such large structures have been extensively studied for offshore engineering issues \citep{Baar,Togn,Muku,Moda2}. VIV of long cables consist of vortex-induced waves, which can be stationary or traveling \citep{Vand2}. The excitation of the structure through lock-in, i.e. the synchronization between the vortex shedding and body oscillation, may occur successively for each vibration mode of the cable \citep{Chap,King}. }

Placing long flexible structures, such as cables, in a cross-flow seems like a promising way to harvest energy from low velocity geophysical flows. Yet the corresponding dynamics is much more complex than that of a rigid body, and it is necessary to explore how the efficiency depends on the parameters of the system. In the present paper, this question is addressed through two reference configurations, using a classical reduced-order model, which is validated here in comparison with Direct Numerical Simulations (DNS) and experiments. The configuration of a tensioned cable with periodically-distributed harvesters is first investigated to study how the distance between harvesters and their respective intensity influence the efficiency. To analyze the impact of a single harvester, and to show the feasibility of the proposed energy harvester in practice, the second configuration is a hanging cable with one harvester at its upper extremity, for which the tension is induced by gravity. Section \ref{sec:model} describes the model used throughout the paper. In Section \ref{sec:rigid}, the case of the elastically-mounted rigid cylinder is considered. Sections \ref{sec:tensioned_cable} and \ref{sec:string} address the two reference cases of a cable with harvesters, introduced above.

\section{Model}	\label{sec:model}

\subsection{A reduced-order wake-oscillator model}

A comprehensive study of the impact on the efficiency of the system parameters using experiments or DNS would be very time consuming and computationally expensive. A reduced-order model is therefore used in this study, based on the ones that have been developed for VIV since \citet{Hart}. These models have been extended to long cables \citep{Viol,Xu,Srin}, and have been proven to predict accurately the main features of their dynamics, as well as some of their complex features like mode switching \citep{Viol2}. In these models, the fluctuating lift exerted by the wake on the bluff body is modeled by a single variable, $q = 2 C_{L}/C_{L0}$, where $C_{L}$ is the instantaneous lift coefficient and $C_{L0}$ the lift coefficient if the solid were fixed. In the wake-oscillator considered here, the evolution of $q$ is assumed to follow a Van der Pol oscillator equation,

\begin{equation}  
	\ddot{q} + \varepsilon \left( q^{2} - 1 \right) \dot{q} + q = f_{s},
	\label{eq:VdP}
\end{equation}

\noindent{where $\dot{\left( \mbox{ } \right)}$ stands for the derivative with respect to the dimensionless time $t = \omega_{f} T$, where $T$ is the time and $\omega_{f} = 2 \pi \mbox{St} U / D$ is the Strouhal shedding frequency, St being the Strouhal number, $U$ the flow speed and $D$ the solid diameter. Equation \eqref{eq:VdP} is coupled with the solid equation by the forcing term $f_{s}$. \citet{Facc} showed that an inertial coupling provides an accurate representation of VIV, using $f_{s} = A \ddot{y}$, where $y = Y/D$ is the solid dimensionless transverse displacement. In the following, the values $\varepsilon = 0.3$ and $A = 12$ are used \citep{Facc}, unless specified otherwise.}

\subsection{Model for the energy harvesters}	\label{subsec:harvesting}

Energy harvesting induces a loss of energy for the fluid-solid system. It is thus represented in the remainder of the paper by a local viscous damping force, whose intensity is a key parameter of the problem.

\section{The elastically-mounted rigid cylinder}	\label{sec:rigid}

\begin{figure}[!t]
	\begin{center}
	%\begin{minipage}{0.3\linewidth}
	%\begin{flushleft}
	%\psfrag{D}[cc][cc][0.75]{$D$}
	%\psfrag{U}[bc][cc][0.8]{$U$}
	%\psfrag{h3}[cc][cc][0.8]{$k$}	
	%\psfrag{xi3}[cc][cc][0.8]{$R$}
	%\psfrag{x}[cc][rc][0.55]{$X$}
	%\psfrag{y}[cc][cc][0.55]{$Y$}
	%\psfrag{z}[cc][cc][0.55]{$Z$}
	%\psfrag{a}[bc][tc][0.75]{(a)}	
	%\includegraphics[width = 0.95\linewidth]{Figure_1a.eps}
	%\end{flushleft}
	%\end{minipage}
	%\begin{minipage}{0.3\linewidth}
	%\begin{center}
	%\psfrag{Xi}[bc][cc][0.8]{$R$}
%	\psfrag{U}[bc][cc][0.8]{$U$}
%	\psfrag{L}[cc][cc][0.8]{$L$}	
%	\psfrag{Theta}[cc][cc][0.8]{$\Theta$}
%	\psfrag{Theta2}[cc][cc][0.8]{$\Theta$}	
%	\psfrag{b}[bc][tc][0.75]{(b)}	
%	\includegraphics[width = 0.85\linewidth]{Figure_1b.eps}
%	\end{center}
%	\end{minipage}
%	\begin{minipage}{0.3\linewidth}
%	\begin{flushright}
%	\psfrag{Xi}[tc][bc][0.8]{$R$}
%	\psfrag{U}[bc][cc][0.8]{$U$}
%	\psfrag{L}[cc][cc][0.8]{$L$}	
%	\psfrag{g}[cc][cc][0.8]{$g$}	
%	\psfrag{c}[bc][tc][0.75]{(c)}	
%	\includegraphics[width = 0.85\linewidth]{Figure_1c.eps}
%	\end{flushright}
%	\end{minipage}
\includegraphics[width=.9\textwidth]{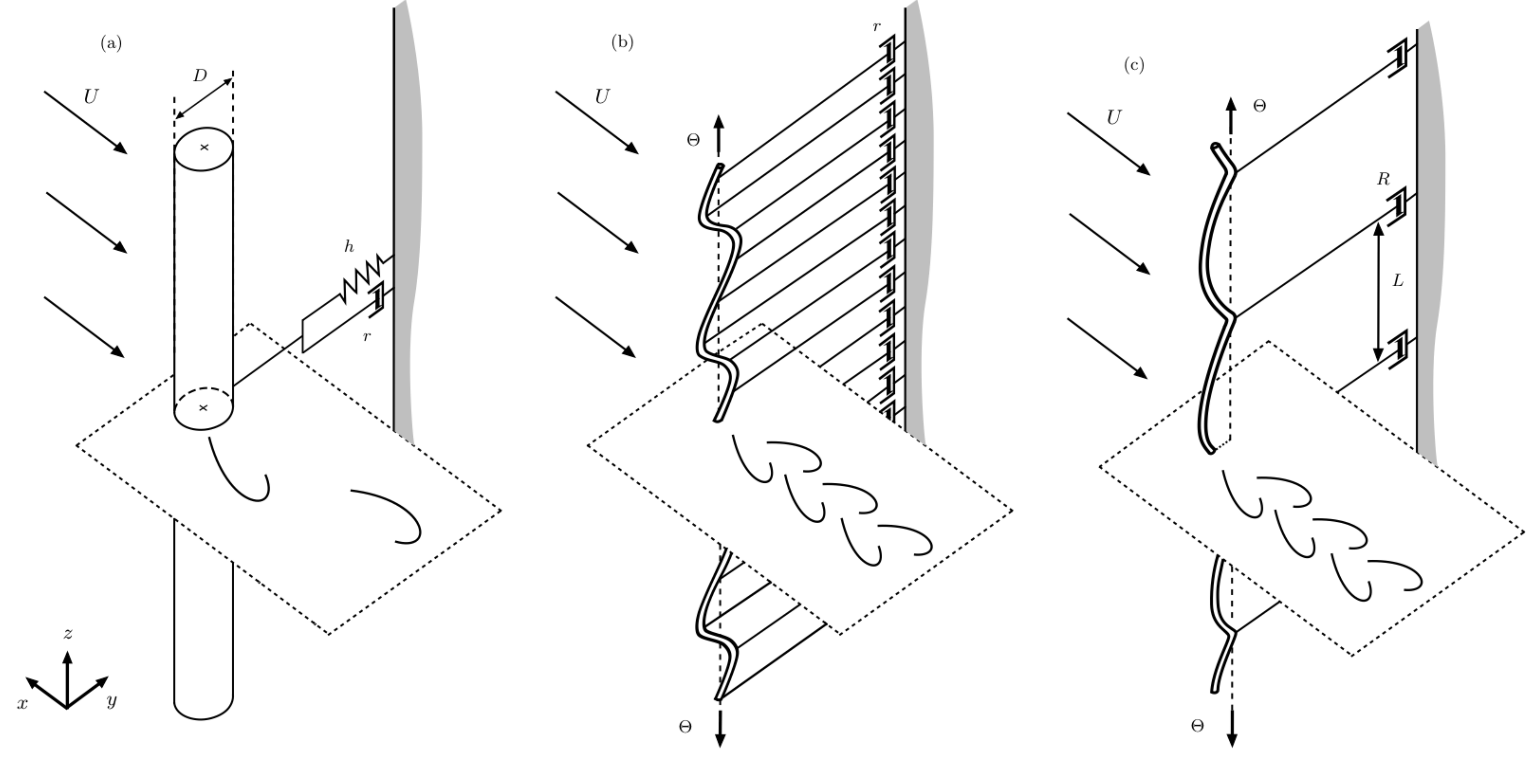}
	\end{center}

	\caption{Energy harvesting by cross-flow VIV of (a) an elastically-mounted rigid cylinder, (b) a tensioned cable with periodically-distributed energy harvesters and (c) a hanging cable with a single energy harvester at its upper extremity.}
\label{fig:scheme} 
\end{figure}

The efficiency of the energy harvesting by VIV of an elastically-mounted rigid cylinder is first investigated,Figure \ref{fig:scheme} (a). Let $m_{s}$ be the mass per unit length of the cylinder. The natural frequency of the cylinder in still fluid may be expressed as $\omega_{s} = \sqrt{k/m_{t}}$, where $m_{t} = m_{s} + \rho \pi D^{2} C_{M0} / 4$ is the total mass per unit length of the cable, including the added inertia. An added mass coefficient of $C_{M0} = 1$ is considered in this paper. Following \citet{Facc}, the cross-flow dynamics of the cylinder is governed by 

\begin{equation}
	\ddot{y} + \left( \xi + \dfrac{\gamma}{\mu} \right) \dot{y} + \delta^{2} y = M q,
	\label{eq:cylinder}
\end{equation}  

\noindent{where $\delta = \omega_{s} / \omega_{f}$ is the ratio between the solid natural frequency and the vortex shedding frequency. The damping term includes both the structural damping $\xi = R/m_{t}\omega_{f}$ and the fluid damping $\gamma / \mu$, where $\gamma = C_{D}/4 \pi St$ is the stall parameter, $C_{D}$ being the drag coefficient \citep{Skop2,Facc}, and $\mu = m_{t} / \rho D^{2}$ is the mass ratio. In Equation \eqref{eq:cylinder}, $M q$ is the wake forcing with $M = C_{L0} / 16 \mu \pi^{2} St^{2}$. Unless specified otherwise, we use $C_{D} = 2$, $C_{L0} = 0.8$, $St = 0.17$, and $\mu = 2.79$ \citep{Viol2}. }

\begin{figure}[!t]
	\begin{center}
%	\psfrag{xi}[cc][cc][1]{$\xi$}
%	\psfrag{delta}[cc][cc][1][-90]{$\delta$}
%	\psfrag{x0}[cc][cc][0.65]{$10^{-2}$}	
%	\psfrag{x1}[cc][cc][0.65]{$10^{-1}$}	
%	\psfrag{x2}[cc][cc][0.65]{$10^{0}$}	
%	\psfrag{x3}[cc][cc][0.65]{$10^{1}$}
%	\psfrag{y1}[cc][cc][0.65]{$1$}	
%	\psfrag{y2}[cc][cc][0.65]{$2$}	
%	\psfrag{y3}[cc][cc][0.65]{$3$}	
%	\psfrag{c1}[lc][cc][0.65]{$0.1$}
%	\psfrag{c2}[lc][cc][0.65]{$0.2$}				
%	\psfrag{b}[tc][bc][0.75]{}	
	\includegraphics[width = 0.5\linewidth]{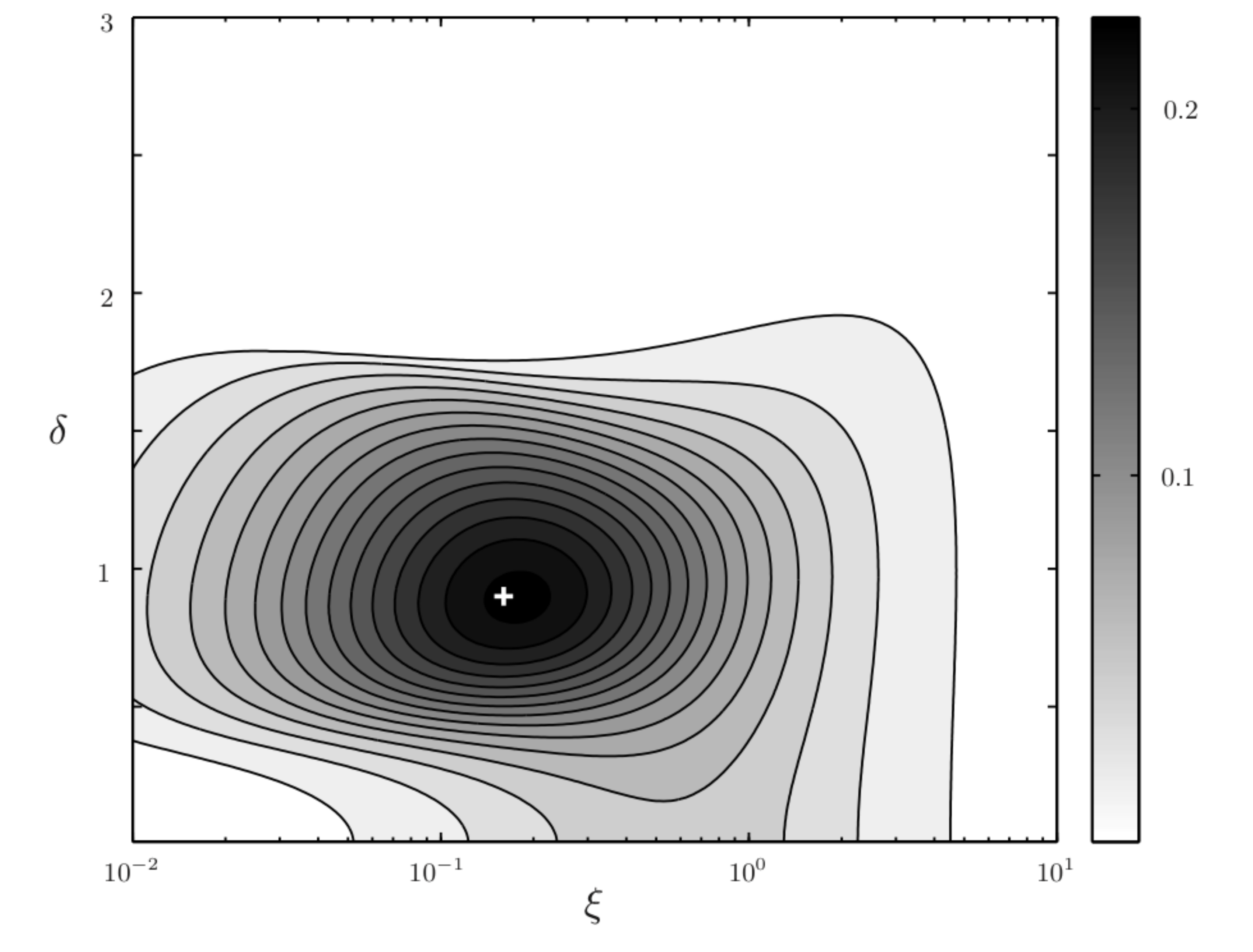}
	\end{center}
	\caption{Efficiency of the energy harvesting by VIV of an elastically-mounted rigid cylinder, as a function of the damping parameter $\xi$ and the frequency ratio $\delta$. The white cross locates the optimal harvesting configuration: $\xi = 0.16$ and $\delta = 0.90$, leading to an efficiency of $\eta = 0.23$.}
\label{fig:rigid} 
\end{figure}

Following Section \ref{subsec:harvesting}, the harvested energy is defined and computed as the total energy dissipated by the structural damping $\xi$. The system of Equations \eqref{eq:VdP} and \eqref{eq:cylinder} is integrated numerically using finite differences and a centered explicit scheme. The solid is initially at rest and the lift coefficient $q$ is subjected to a small random initial perturbation. The system quickly reaches a limit cycle which is used to derive time-averaged quantities such as the efficiency, Equation \eqref{eq:efficiency}, which reads as $\eta = 16 \mu \pi^{3} St^{3} \left< \xi \dot{y}^{2}\right>$ in dimensionless variables. The evolution of $\eta$ with the damping $\xi$ and the frequency ratio $\delta$ is shown inFigure \ref{fig:rigid}. The maximum efficiency, $\eta_{opt} = 0.23$, is close to the value of 0.22 presented in \citet{Bern}. It is obtained under lock-in, which results in high-amplitude vibrations. More precisely, the optimal efficiency is reached for $\xi_{opt} = 0.16$ and $\delta_{opt} = 0.90$, in the vicinity of the condition, $\delta = 1$. For each frequency ratio $\delta$, the efficiency follows a bell-shaped evolution with $\xi$. At low damping parameters, $\eta$ is proportional to $\xi$ as VIV amplitude saturates \citep{Will}. The amplitude is proportional to $1/\xi$ at high damping parameters \citep{Facc}, and so is the efficiency. 

We have shown here that the wake oscillator model is a valuable tool to simply derive efficiency maps. Moreover we have emphasized that high efficiencies are obtained under lock-in condition, in a region of the parameter space where the frequency of shedding matches the natural frequency of the cylinder. The successive lock-ins of the different vibration modes of a flexible structure may consequently lead to wider ranges of parameters leading to high efficiencies. Energy harvesting by VIV of cables is therefore explored in the next sections.

\section{A tensioned cable with periodically-distributed harvesters}	\label{sec:tensioned_cable}

In this section, the energy harvesting by cross-flow VIV of an infinitely long tensioned cable with periodically-distributed energy harvesting devices is investigated,Figure \ref{fig:scheme} (b). This discussion does not aim at describing how to implement theses devices, as obviously infinitely long cables cannot be constructed. The discussion, however, is valuable towards understanding the impact  on the efficiency of the harvester distribution along a cable. The parameters $D$, $m_{s}$,  $\rho$ and $U$ have the same meaning as in Section \ref{sec:rigid}, and $\Theta$ is now the cable's uniform tension. The energy harvesting devices are modeled by dashpots of damping coefficient $R$ and the distance between two successive dashpots is denoted by $L$. The dimensionless spanwise coordinate is defined as $z = Z/Z_{c}$, with $Z_{c}$ the wavelength of a wave traveling along the cable with frequency $\omega_{f}$ and phase speed $c = \sqrt{\Theta/m_{t}}$,

\begin{equation}
	Z_{c} = \dfrac{ 2 \pi}{\omega_{f}} \sqrt{\dfrac{\Theta}{m_{t}}}.
	\label{eq:Zc_tension}
\end{equation}

\noindent{The cross-flow dynamics of the cable are described by the dimensionless equation}

\begin{equation}
	\ddot{y} + \dfrac{\gamma}{\mu} \dot{y} - \dfrac{1}{4 \pi^{2}} y^{\prime \prime} = M q,
	\label{eq:eq_cable}
\end{equation}

\noindent{where $\left( \mbox{ } \right)^{\prime}$ stands for the derivation with respect to $z$. The force balance at the dashpot locations introduces the local condition}

\begin{equation}
	\dfrac{1}{4 \pi^{2}} \left[ y^{\prime} \left( 0,t \right) - y^{\prime} \left( \ell, t \right) \right] = \xi \ell \dot{y} \left( 0,t \right),
	\label{eq:BC_cable}
\end{equation}

\noindent{where $y^{\prime} \left( 0,t \right)$ is the right derivative of the displacement $y$ with respect to $z$ at location $z=0$, and $y^{\prime} \left( \ell, t \right)$ its left derivative at $z=\ell$. The damping parameter reads as $\xi = R / L m_{t} \omega_{f}$, and the reduced length is defined as $\ell = L/Z_{c}$. The motion of the cable with periodic harvesters, Equations \eqref{eq:VdP}, \eqref{eq:eq_cable} and \eqref{eq:BC_cable}, is integrated using centered finite differences on a spatially periodic domain and a Crank-Nicolson method.}

\subsection{Comparison with direct numerical simulations}	\label{subsec:DNS}

\begin{figure}[!ht]
	\begin{center}
\includegraphics[width=.95\textwidth]{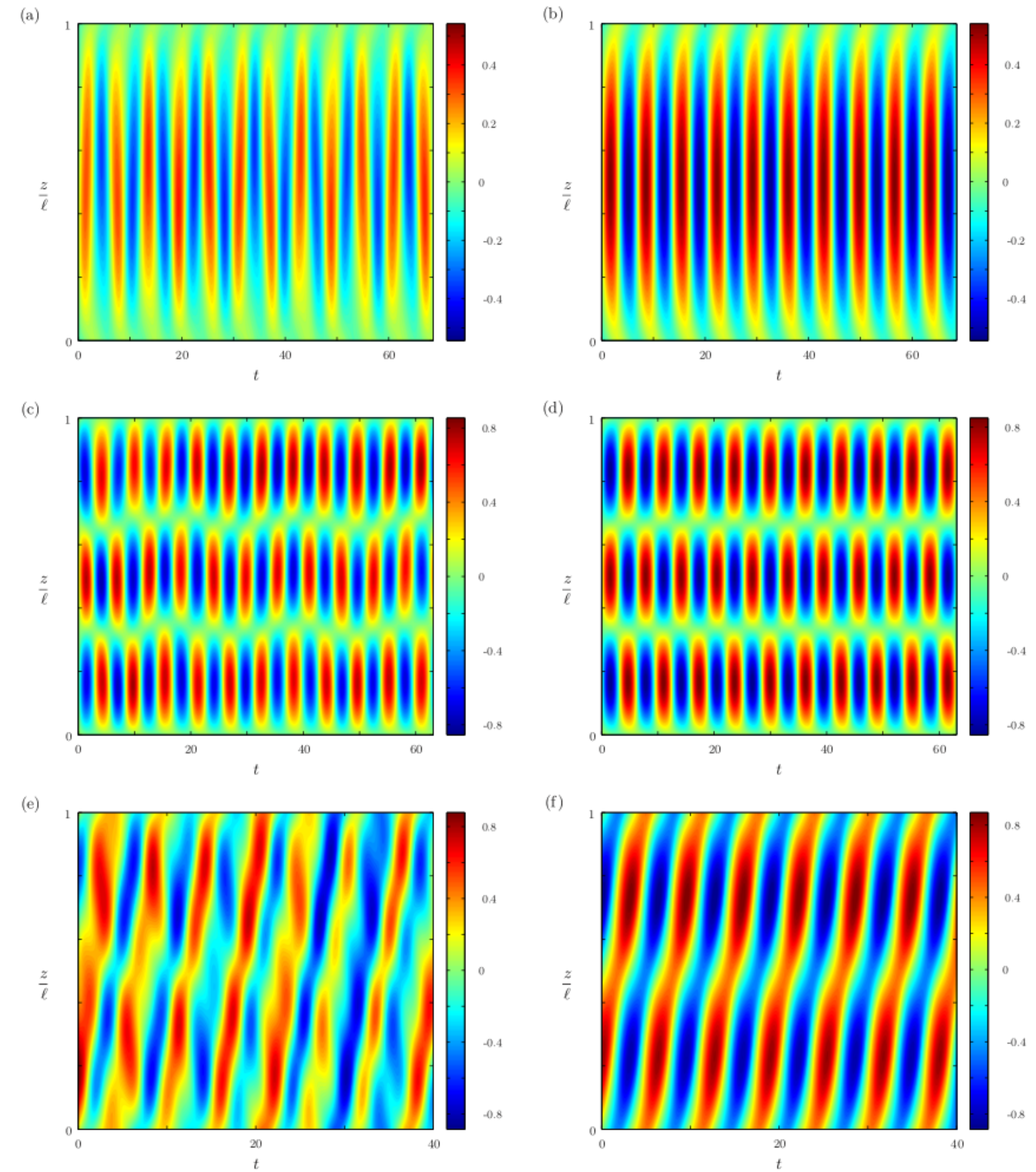}
	\end{center}
	\caption{Cross-flow displacement of the tensioned cable predicted by DNS (left) and by the model (right): $ \ell = 0.55$ and $\xi = 3.65$ inFigures (a) and (b), $\ell = 1.50$ and $\xi = 3.65$ inFigures (c) and (d), $\ell = 1.00$ and $\xi = 0.04$ inFigures (e) and (f).}
	\label{fig:comp_dynamics}
\end{figure}

The cable dynamics predicted by the model is first compared with the results of direct numerical simulations (DNS),Figure \ref{fig:comp_dynamics}. The parallelized code Nektar, based on the spectral/hp element method \citep{Karn} is used to solve the coupled fluid-solid dynamics, as detailed in \citet{Bour}. The Reynolds number is set to $\mbox{Re} = 300$ and the corresponding lift coefficient $C_{L0}$ is obtained by a DNS with a fixed cable. The obtained value of $C_{L0} = 0.61$ is consistent with the experimental results found in \citet{Norb} and is used in this section. 

\begin{figure}[!t]
	\begin{center}
%		\begin{minipage}{0.45\linewidth}
%			\psfrag{1}[cc][cc][0.8]{}
%			\psfrag{x0}[tc][cc][0.65]{0}
%			\psfrag{x1}[tc][cc][0.65]{1}
%			\psfrag{x2}[tc][cc][0.65]{2}
%			\psfrag{x3}[tc][cc][0.65]{3}
%			\psfrag{y0}[cc][cc][0.65]{0}
%			\psfrag{y02}[rc][cc][0.65]{0.02}
%			\psfrag{y04}[rc][cc][0.65]{0.04}
%			\psfrag{y06}[rc][cc][0.65]{0.06}
%			\psfrag{y08}[rc][cc][0.65]{0.08}
%			\psfrag{y1}[rc][cc][0.65]{0.1}
%			\psfrag{l}[tc][bc][0.85]{$\ell$}
%			\psfrag{Eff}[rc][lc][0.85][-90]{$\eta$}
%			\psfrag{a}[cc][cc][0.65]{(a)}
%			\psfrag{c}[cc][cc][0.65]{(c)}			
%			\includegraphics[width = 0.92\linewidth]{Figure_4a.eps}
%		\end{minipage}	\hspace{0.5cm}
%		\begin{minipage}{0.45\linewidth}
%			\psfrag{2}[cc][cc][0.8]{}
%			\psfrag{xm2}[tc][cc][0.65]{$\mbox{10}^{-2}$}
%			\psfrag{xm1}[tc][cc][0.65]{$\mbox{10}^{-1}$}
%			\psfrag{x0}[tc][cc][0.65]{1}
%			\psfrag{x1}[tc][cc][0.65]{10}
%			\psfrag{x2}[tc][cc][0.65]{100}
%			\psfrag{y0}[cc][cc][0.65]{0}
%			\psfrag{y01}[rc][cc][0.65]{0.01}
%			\psfrag{y02}[rc][cc][0.65]{0.02}
%			\psfrag{y03}[rc][cc][0.65]{0.03}
%			\psfrag{y04}[rc][cc][0.65]{0.04}
%			\psfrag{Xi}[tc][bc][0.85]{$\xi$}
%			\psfrag{Eff}[rc][lc][0.85][-90]{$\eta$}
%			\psfrag{e}[cc][cc][0.65]{(e)}
%			\includegraphics[width = 0.92\linewidth]{Figure_4b.eps}
%		\end{minipage}
\includegraphics[width=.9\textwidth]{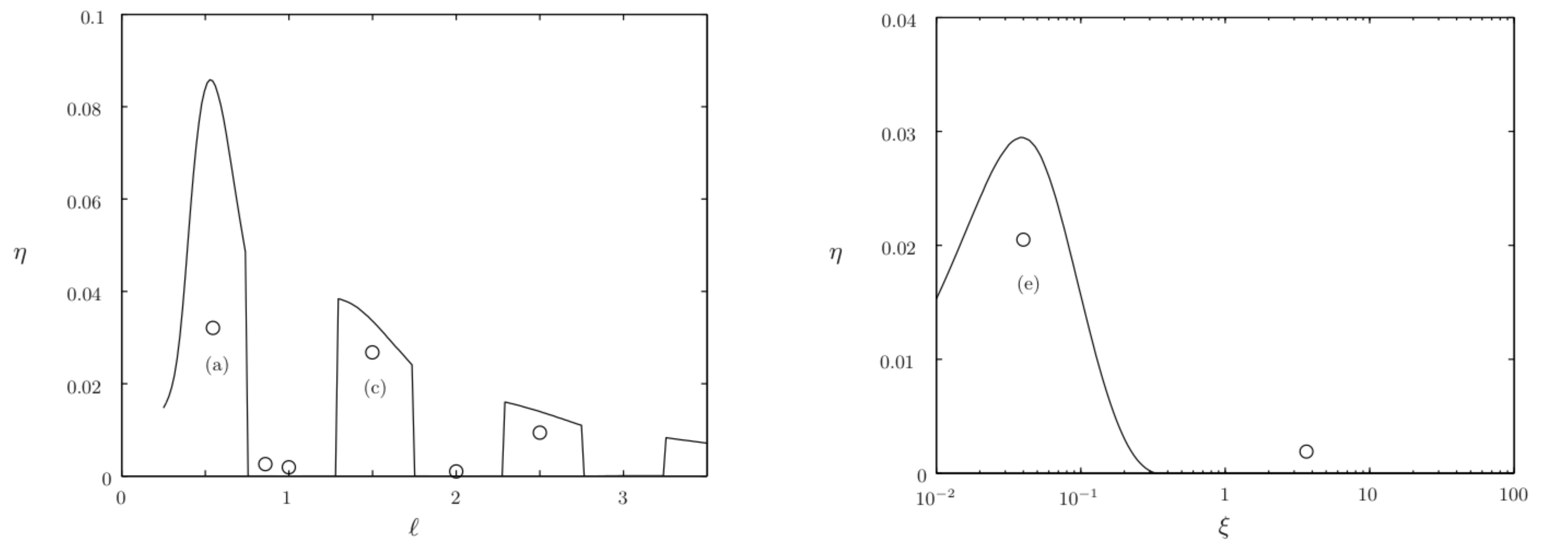}
	\end{center}
	\caption{Left: evolution of the efficiency with the reduced length $\ell$ for $\xi = 3.65$; Right: evolution of the efficiency with the damping parameter $\xi$ for $\ell = 1$. Model results (solid lines) and DNS results (symbols). The letters (a), (c) and (e) correspond to the DNS results shown inFigure \ref{fig:comp_dynamics}.}
	\label{fig:local_comparison}
\end{figure}

The displacements predicted by the model and the DNS are plotted inFigure \ref{fig:comp_dynamics}), for three different sets of parameters. In both cases, the vibration modes are identical. A good agreement is also observed between the model and the DNS for predicting the traveling or stationary behaviour of the cable VIV.Figure \ref{fig:local_comparison} shows the evolution of the efficiency $\eta = 16 \mu \pi^{3} St^{3} \left< \xi \dot{y}^{2}\right>$ with the two harvesting parameters, $\ell$ and $\xi$. For both DNS results and model predictions, the efficiency $\eta$ varies rather smoothly with the damping parameter $\xi$, while its evolution with $\ell$ exhibits some discontinuities. This strong influence of the reduced length on the efficiency will be discussed in section \ref{subsec:Efficiency_cable}. These comparisons between DNS and model predictions confirm the ability of the model to represent the dynamics of cross-flow VIV of a tensioned cable. The VIV model is consequently an appropriate tool to complete a systematic study of the evolution of the efficiency $\eta$ with the two harvesting parameters, $\xi$ and $\ell$. 

\subsection{Efficiency map}	\label{subsec:Efficiency_cable}

\begin{figure}[!t]
%		\begin{center}
%		\psfrag{x0}[cc][cc][0.75]{$\mbox{10}^{-2}$}
%		\psfrag{x1}[cc][cc][0.75]{$\mbox{10}^{-1}$}
%		\psfrag{x2}[cc][cc][0.75]{$\mbox{10}^{0}$}
%		\psfrag{x3}[cc][cc][0.75]{$\mbox{10}^{1}$}	
%		\psfrag{y1}[cc][cc][0.875]{1}
%		\psfrag{y2}[cc][cc][0.75]{2}			
%		\psfrag{y3}[cc][cc][0.75]{3}	
%		\psfrag{c1}[lc][cc][0.75]{0}			
%		\psfrag{c2}[lc][cc][0.75]{0.1}	
%		\psfrag{c3}[lc][cc][0.75]{0.2}	
% 		\psfrag{01}[cc][cc][0.8]{0 $\rightarrow$ 1}			
% 		\psfrag{2}[cc][cc][0.8]{2}	
% 		\psfrag{3}[cc][cc][0.8]{3}	
% 		\psfrag{4}[cc][cc][0.8]{4}	
% 		\psfrag{5}[cc][cc][0.8]{5}
% 		\psfrag{6}[cc][cc][0.8]{6}
% 		\psfrag{7}[cc][cc][0.8]{7}
% 		\psfrag{8}[cc][cc][0.8]{8}									
%		\psfrag{l}[cc][cc][1][-90]{$\ell$}	
%		\psfrag{Xi}[cc][cc][1]{$\xi$}
%	 	\includegraphics[width = 0.7\linewidth]{Figure_5.eps}
%		\end{center}
\includegraphics[width=.7\textwidth]{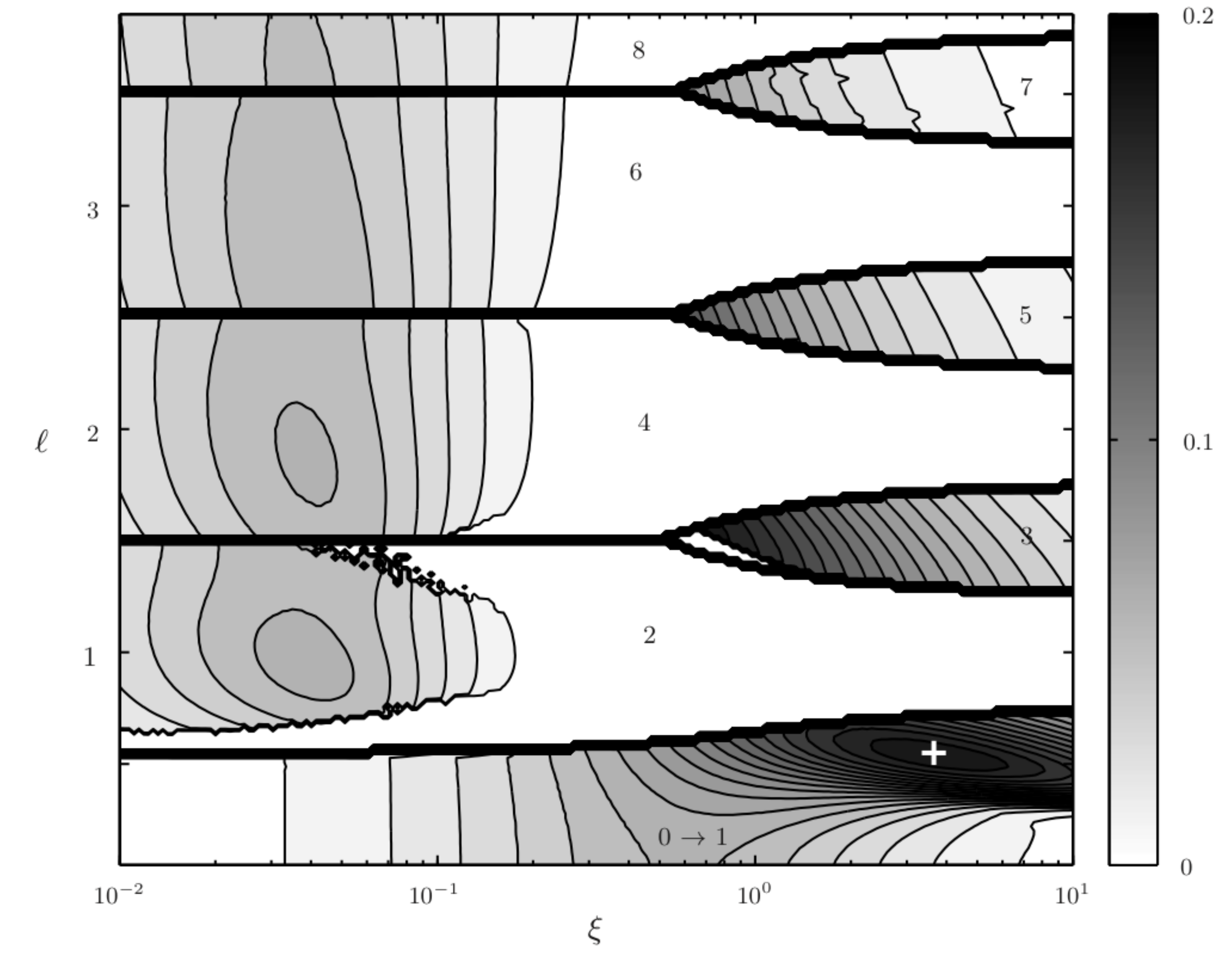}
		\caption{Efficiency $\eta$ as a function of $\xi$ and $\ell$. The white cross locates the optimal harvesting configuration: $\xi = 3.65$ and $\ell = 0.55$, leading to an efficiency of $\eta = 0.19$. The number of the dominant mode, determined by a linear stability analysis, is also indicated and the heavy lines mark the boundaries between the dominance zones of each mode number.}
		\label{fig:Map_cable}
	\end{figure}

A full parametric study of the efficiency $\eta$ is performed, with the same parameters as in Section \ref{sec:rigid}, in particular $C_{L0} = 0.8$ contrary to the previous section for which it was adapted to the considered values of the Reynolds number.Figure \ref{fig:Map_cable} shows the evolution of $\eta$ with respect to the reduced length $\ell$ and the damping parameter $\xi$.

The optimal efficiency $\eta_{opt} = 0.19$, reached for $\ell_{opt} = 0.55$ and $\xi_{opt} = 3.65$, is close to the value of 0.23 obtained for a rigid cylinder, Section \ref{sec:rigid}. The efficiency $\eta$ depends significantly on both harvesting parameters $\ell$ and $\xi$. Except for $\ell < 0.5$, where its variations are rather smooth, the efficiency varies dramatically with the reduced length $\ell$. It exhibits discontinuities, which have already been pointed out inFigure \ref{fig:local_comparison} (left), where the high impact of $\ell$ on $\eta$ is visible. The high-efficiency zones correspond to well-defined high efficiency tongues located in the high-damping region of the map; these regions are analyzed in the following.  

\subsection{Relation between efficiency and vibration modes}

\begin{figure}[!t]
		\begin{center}
\includegraphics[width=.9\textwidth]{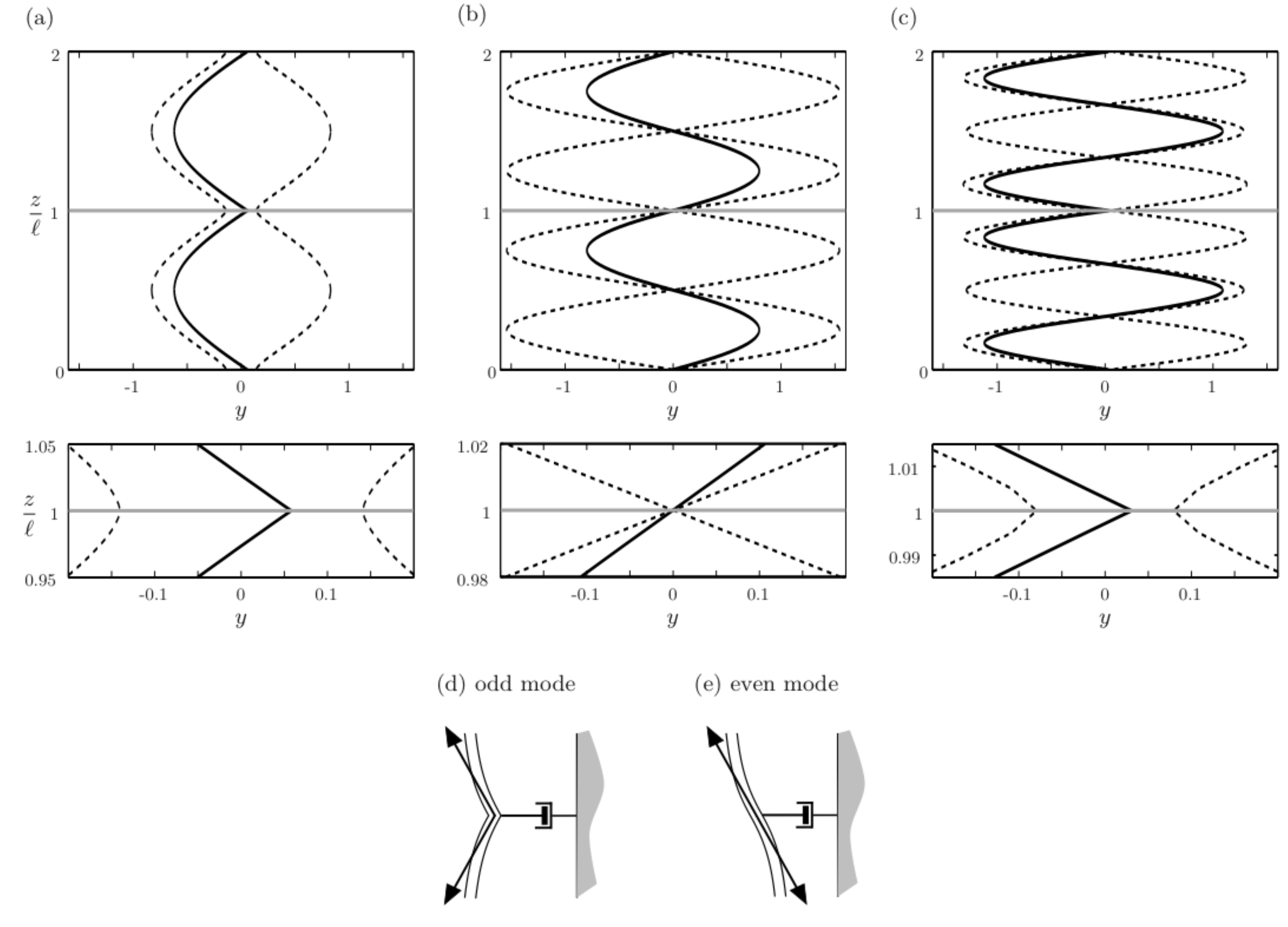}
\end{center}
		\caption{Displacement (solid line) and envelope (dashed line) of the cross-flow VIV of the tensioned cable for $\xi = \xi_{opt} = 3.65$ and (a) $\ell = \ell_{opt} = 0.55$, corresponding to the optimal case for energy harvesting, $\eta = \eta_{opt} = 0.19$ (b) $\ell = 1$, $ \eta = 0$ and (c) $\ell = 1.50$, $\eta = 0.07$. A magnification of each plot is shown belowFigures (a), (b) and (c) in order to visualize the displacement of the cable in the vicinity of the dashpots, $z/\ell \approx 1$. In eachFigure, the horizontal grey line shows the location of one energy harvester. Sketches of the resulting forcing of the cable on the dashpot are also shown, in cases where the excited mode number is (d) odd and (e) even.}
		\label{fig:comp_modes}
\end{figure}

The cross-flow motion of the cable for three sets of parameters leading to very different efficiencies are shown inFigure \ref{fig:comp_modes}. In order to characterize the cable dynamics, the mode number $n$ is defined as 

\begin{equation}
	n = \dfrac{ 2 \ell}{\lambda},
	\label{eq:def_lambda}
\end{equation}

\noindent{where $\lambda$ is the wavelength of the VIV. As the extremities of the cable are not fixed but damped, this mode number may not necessarily be an integer. In such cases, $n$ is defined as the closest integer to the ratio $2 \ell / \lambda$.}

When the harvesting parameters are optimal, the cable displacement corresponds to a mode 1,Figure \ref{fig:comp_modes} (a), and to a mode 3 when $\ell = 1.50$,Figure \ref{fig:comp_modes} (c). In both cases, the envelope of the oscillations also shows that the dashpots are moving, leading to an efficient extraction of energy. On the contrary, the energy harvesters are still when $\ell = 1$, for which the selected vibration mode has a mode number of $n = 2$,Figure \ref{fig:comp_modes} (b).

The mode selected by the cable VIV therefore seems to be a key parameter for the system efficiency. The dashpots are forced by the slope discontinuity between the two strands of the cable on either side of the harvesters, Equation \eqref{eq:BC_cable}. This slope discontinuity is much larger when the mode number $n$ is odd compared with when it is even, as shown inFigure \ref{fig:comp_modes} (d) and (e). This explains why the parameters for which an odd mode of vibration is excited lead to a much higher efficiency than the ones leading to an even mode. It is consequently important to identify the dominance zone of every vibration mode of the tensioned cable. To this end, a linear approach is used.

\subsection{Linear stability analysis and lock-in}

Following the work of \citet{Lang}, \citet{Viol2} have shown that a linear stability analysis of the VIV model provides an accurate prediction of the frequency and wavelength of VIV of a tensioned cable. The non-linear term $\varepsilon q^{2} \dot{q}$ of the wake-oscillator, Equation \eqref{eq:VdP}, is neglected and a stability analysis of the coupled linear system is performed. The most unstable linear mode is determined for each set of parameters $\ell$ and $\xi$. The dominance zones of the cable vibration modes are reported on the efficiency map,Figure \ref{fig:Map_cable}. There is a very good agreement between the dominance zones of the linear vibration modes and the complex structure of the efficiency map. The high efficiency regions of the parameter space correspond rigorously to the dominance zones of the odd vibration modes. This confirms that in practice, the harvesting parameters $\ell$ and $\xi$ should be chosen so that an odd mode number of the cable is excited, for the harvesting to be efficient. As for the case of an elastically-mounted rigid cylinder, Section \ref{sec:rigid}, the high efficiencies are obtained under lock-in conditions, but for any of the odd modes.

The structure of the efficiency map is therefore due to the dominance zones of the different vibration modes of the cable in VIV. The eigenfrequency of the mode $n$ of a tensioned cable with fixed ends reads as

\begin{equation}
	\omega_{n} = \dfrac{n \pi}{L} \sqrt{\dfrac{\Theta}{m_{t}}}.
	\label{eq:def_wn}
\end{equation}

\noindent{Combining Equation \eqref{eq:def_wn} with the definition of $\ell$, Equation \eqref{eq:Zc_tension}, leads to}

\begin{equation}
	\ell = \dfrac{n}{2} \left( \dfrac{\omega_{f}}{\omega_{n}} \right).
	\label{eq:def_l_lock-in}
\end{equation} 

\noindent{The dimensionless length parameter, $\ell$, corresponds to the ratio between the Strouhal frequency and the successive eigenfrequencies of the tensioned cable. A match between the vortex shedding frequency, $\omega_{f}$, and the natural frequency of mode $n$, $\omega_{n}$, is therefore observed for $\ell = n/2$. These conditions notably locate the local maxima of $\eta$ within the dominance zone of each mode,Figure \ref{fig:Map_cable}. The two harvesting parameters have to be chosen so that $\ell$ ensures that the shedding frequency is close to the natural frequency of an odd mode while $\xi$ has to be high enough to allow this particular mode to be excited; at low damping, only even modes develop, as seen inFigure \ref{fig:Map_cable}.}

The existence of several vibration modes is one of the fundamental characteristics of a flexible structure when compared to an elastically-mounted rigid cylinder. The cable adapts its VIV dynamics to the wake forcing, through mode selection, which strongly impacts the efficiency of the harvesting,Figure \ref{fig:Map_cable}. Therefore, flexibility modifies the energy harvesting behaviour of the structure dramatically, compared to a rigid one, through this ability to adapt its dynamics via the lock-in of its successive modes.

\section{A hanging cable with a localized harvester}	\label{sec:string}

The previous section has shown that VIV of long and flexible structures may be a relevant mechanism to harvest energy from geophysical flows. It is yet still necessary to look for an appropriate configuration with fewer harvesters. In this section, a single energy harvester is considered. The cable is hung by an energy harvester at its upper extremity, the lower one being free,Figure \ref{fig:scheme} (c). This configuration differs from the previous one by its gravity-induced tension, varying along the span of the cable, and by the asymmetry of its boundary conditions.

The dimensionless cross-flow displacement $y$ and time $t$ are defined as in previous sections. Following \citet{Grou}, the dimensionless spanwise coordinate is defined as $z = Z / Z_{c}$ where the characteristic length for the spanwise coordinate is $Z_{c} = m_{s} g / m_{t} \omega_{f}^{2}$. Given these characteristic dimensions, there are two dimensionless harvesting parameters. The first one is $\ell = L / Z_{c}$, the dimensionless length of the hanging cable. The second one is the dimensionless damping parameter, $\xi = R / \sqrt{ L m_{s} m_{t} g}$. 

\subsection{Experiments}	\label{subsec:hanging_manip}

\begin{figure}[t]
	\begin{center}
%	\psfrag{eau}[lc][rc][0.85]{water/glycerol}
%	\psfrag{support}[rc][cc][0.85]{support}
%	\psfrag{laser}[cc][lc][0.85]{laser}
%	\psfrag{U}[bc][cc][0.85]{$U$}
%	\psfrag{L}[bc][cc][0.85]{$L$}	
%	\psfrag{X}[lc][rc][0.6]{$X$}
%	\psfrag{Y}[cc][lc][0.6]{$Y$}
%	\psfrag{Z}[cc][tc][0.6]{$Z$}
%	\psfrag{g}[cc][cc][0.85]{$g$}	
%	\includegraphics[width = 0.35\linewidth]{Figure_7.eps}
\includegraphics[width=.35\textwidth]{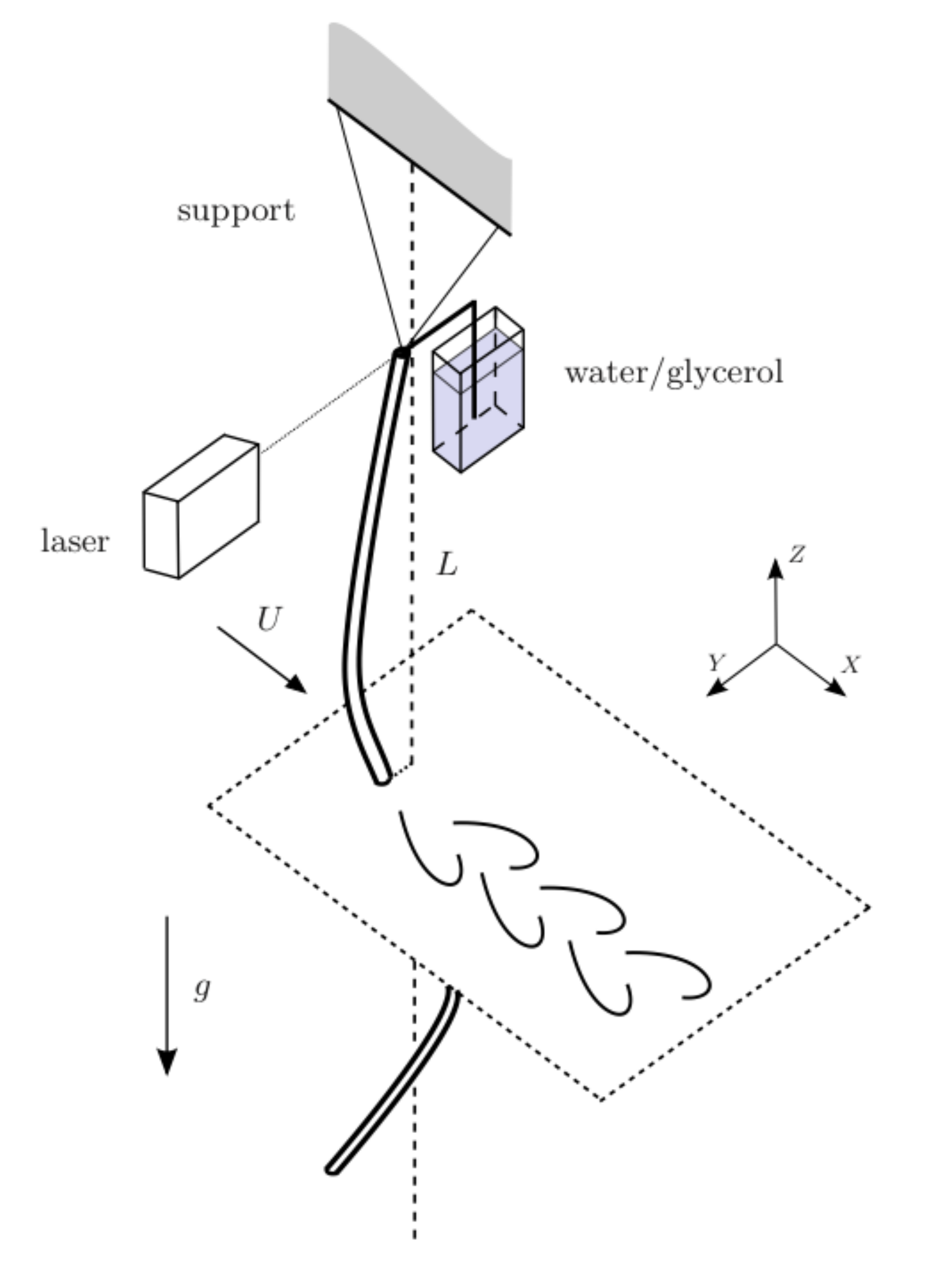}
	\end{center}
	\caption{Sketch of the experimental setup for the analysis of the energy harvesting by cross-flow VIV of a hanging cable in a uniform flow.}
\label{fig:schema_harvesting} 
\end{figure}	

The energy harvesting by VIV of a hanging cable was first studied experimentally; then the results were compared with the predictions of an appropriate model. Physical properties of the cable were $L = 0.65 \mbox{ m}$, $m_{s} = 21 \mbox{ g.m}^{-1}$ and $D = 2.5 \mbox{ mm}$. The cable was placed in uniform flow of constant speed $U$ in the water channel of the Fluid-Structure Interactions laboratory of University of Massachusetts, Amherst, whose test section dimensions are 38 cm $\times$ 50 cm $\times$ 150 cm. The experimental setup was similar to that used by \citet{Grou}, except for the top dissipative boundary condition. The cable was hung by nylon strings and fixed to the dashpot,Figure \ref{fig:schema_harvesting}. The viscous damping was generated by a stick, immersed in an appropriate mixture of water and glycerol. This allowed adjusting the magnitude of the damping parameter by varying the proportion of glycerol. Most of the uncertainty in the forthcoming results actually came from the uncertainty in the measurement of the damping parameter $\xi$, obtained through water-glycerol mixture. The displacement of the upper end of the cable was measured with a laser displacement sensor Micro-epsilon OptoNCDT ILD1302-20. The acquisition frequency was 256 Hz. Three flow speeds were tested, corresponding to different values of the dimensionless length parameter: $\ell = 15.0$, $34.1$ and $50.0$. For each flow speed, measurements were taken for five damping values ranging from 0.1 to 20. The measured displacement was filtered around its dominant frequency, close to the expected Strouhal frequency \citep{Grou}. To avoid noise due to signal differentiation, the numerator of Equation \eqref{eq:efficiency} was estimated as $R \Omega^{2} Y_{L}^{2}$, with $\Omega$ and $Y_{L}$ the frequency and amplitude of the cable upper end oscillation. 

\begin{figure}[!ht]
	\begin{center}
%	\psfrag{y0}[bc][lc][0.65]{$0$}
%	\psfrag{y002}[cc][lc][0.65]{$0.002$}
%	\psfrag{y004}[cc][lc][0.65]{$0.004$}
%	\psfrag{y006}[cc][lc][0.65]{$0.006$}
%	\psfrag{x2}[cc][cc][0.65]{$10^{2}$}			
%	\psfrag{xm2}[cc][cc][0.65]{$10^{-2}$}	
%	\psfrag{xm1}[cc][cc][0.65]{$10^{-1}$}	
%	\psfrag{x0}[cc][cc][0.65]{$1$}	
%	\psfrag{x1}[cc][cc][0.65]{$10$}	
%	\psfrag{xi}[tc][bc][0.85]{}	
%	\psfrag{a}[bc][tl][0.85]{(a)}	
%	\psfrag{b}[bc][tl][0.85]{(b)}	
%	\psfrag{c}[bc][tl][0.85]{(c)}	
%	\psfrag{eff}[tc][bc][0.85][-90]{$\eta$}	
%	\includegraphics[width = 0.5\linewidth]{Figure_8a.eps} \\
%	\includegraphics[width = 0.5\linewidth]{Figure_8b.eps} \\
%	\psfrag{xi}[tc][bc][0.85]{$\xi$}	
%	\includegraphics[width = 0.5\linewidth]{Figure_8c.eps} 
\includegraphics[width=.45\textwidth]{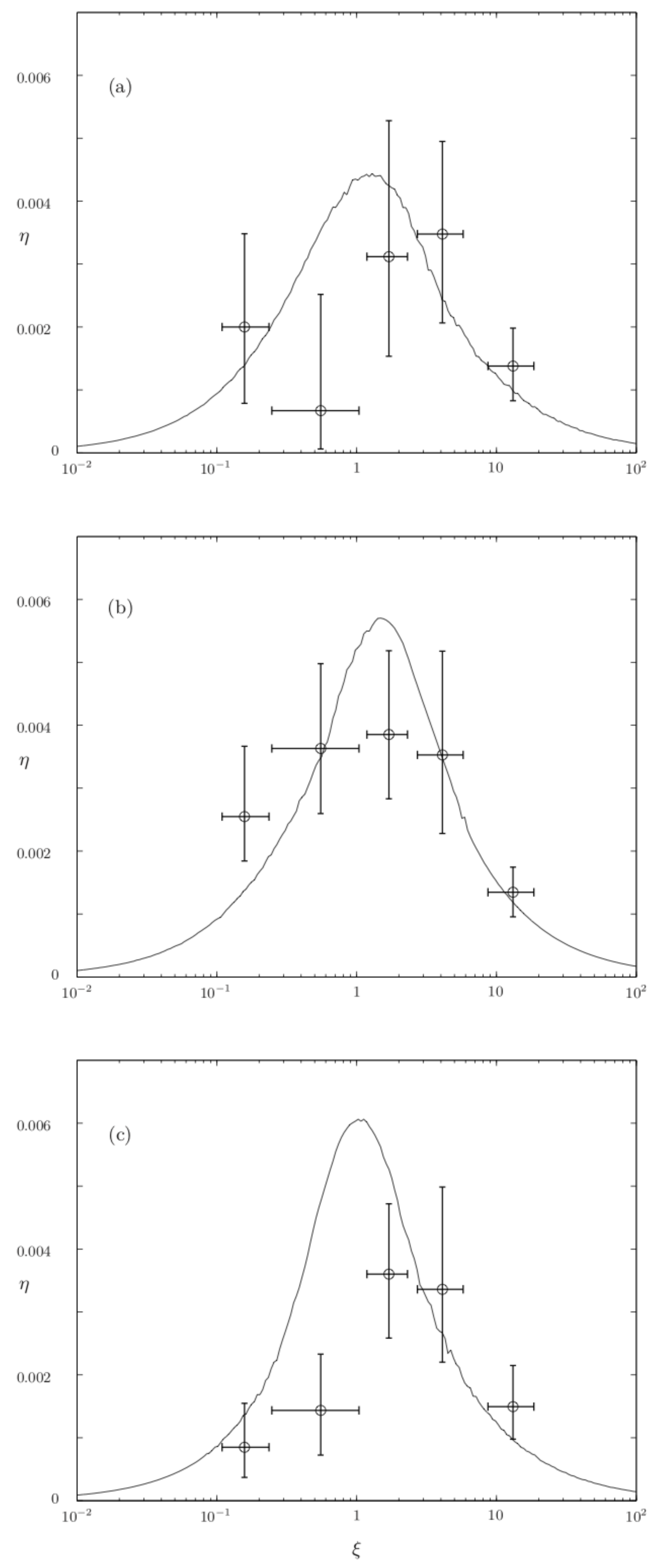}
	\end{center}
	\caption{Efficiency $\eta$ obtained experimentally (symbols) and by the model for VIV of a hanging string (solid line) for (a) $\ell = 15.0$, (b) $\ell = 34.1$ and (c) $\ell = 50.0$.}
\label{fig:Eff_WO_manip} 
\end{figure}

Figure \ref{fig:Eff_WO_manip} shows the evolution of the efficiency with the damping parameter $\xi$, derived from the experiments. For all values of $\ell$, the efficiency shows a maximum, in the vicinity of $\xi = 1$. This configuration is also studied using the wake oscillator model coupled with a cable equation, as in the previous section, but taking into account the gravity-induced spatial variations of the tension and the specific boundary conditions. The model reads

\begin{subeqnarray}
	 \ddot{y} + \frac{\gamma}{\mu} \dot{y} - \left( z y^{\prime} \right)^{\prime} & = & M q, 	\slabel{eq:WO_corde_eff1} \\
	 \ddot{q} + \varepsilon \left( q^{2} - 1 \right) \dot{q} + q & = & A \ddot{y},	\slabel{eq:WO_corde_eff2}
	 \label{eq:WO_corde_eff}
\end{subeqnarray}	

\noindent{with boundary conditions}

\begin{equation}
	m_{s} \ddot{y} - m_{t} y^{\prime} = 0 \quad \mbox{ and } \quad y^{\prime} - \dfrac{y}{h} + \dfrac{\xi}{\sqrt{\ell}} \dot{y} = 0,
	\label{eq:BC_manips}
\end{equation}

\noindent{at the lower end, $z=0$, and upper end, $z = \ell$, respectively. The lower end condition is obtained from the momentum balance for an arbitrary small end mass, and the upper end condition corresponds to a connection between a massless nylon string of length $h$ and the damper. The lift coefficient $C_{L0}$ is adapted to the experimental range of Reynolds numbers, $\mbox{Re} \approx 100$, $C_{L0} = 0.3$ \citep{Norb,Viol}. The coupling parameter $A$ is fitted to amplitude measurements of the hanging cable VIV in cases were the upper end is fixed, $A = 2.5$ \citep{Grou2}, while the Strouhal number is taken from \citet{Grou}. The model is integrated in space and time using finite differences and the efficiency reads as}

\begin{equation}
	\eta = \dfrac{16 \mu \pi^{3} St^{3} \left< \xi \dot{y}^{2}\right>}{ \sqrt{\ell}}.
	\label{eq:def_eff_corde}
\end{equation}

\noindent{The evolution of the efficiency, $\eta$, with $\xi$ is plotted inFigure \ref{fig:Eff_WO_manip}. There is a reasonable agreement between the experimental results and numerical predictions. Computational results clearly show the expected bell-shaped evolution of the efficiency, consistently with the experimental results. Note that the length parameter $\ell$ has a rather small influence on the results, contrary to the previous configuration.}

\subsection{Parametric study}	\label{subsec:eff_corde}

\begin{figure}[t]
		\begin{center}
%		\psfrag{x0}[cc][cc][0.75]{$\mbox{10}^{-2}$}
%		\psfrag{x1}[cc][cc][0.75]{$\mbox{10}^{-1}$}
%		\psfrag{x2}[cc][cc][0.75]{$\mbox{10}^{0}$}
%		\psfrag{x3}[cc][cc][0.75]{$\mbox{10}^{1}$}
%		\psfrag{x4}[cc][cc][0.75]{$\mbox{10}^{2}$}	
%		\psfrag{y5}[cc][cc][0.75]{50}			
%		\psfrag{y10}[cc][cc][0.75]{100}
%		\psfrag{y15}[cc][cc][0.75]{150}
%		\psfrag{y20}[cc][cc][0.75]{200}		
%		\psfrag{c1}[lc][cc][0.75]{0}			
%		\psfrag{c2}[lc][cc][0.75]{0.1}	
%		\psfrag{c3}[lc][cc][0.75]{0.2}						
%		\psfrag{ell}[cc][cc][0.9][-90]{$\ell$}	
%		\psfrag{Xi}[tc][bc][0.9]{$\xi$}			
		\includegraphics[width = 0.7\linewidth]{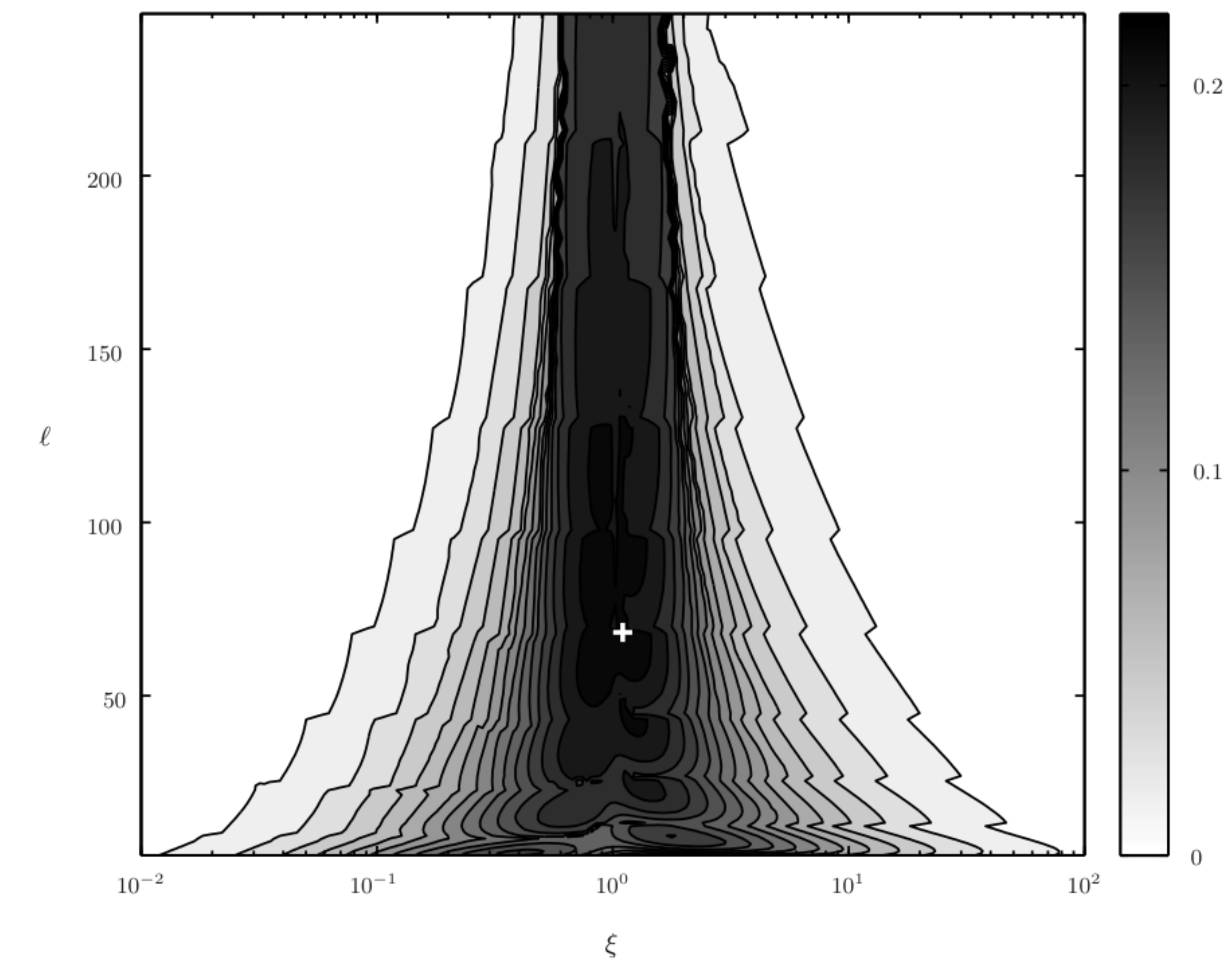}
		\end{center}
		\caption{Efficiency of the energy harvesting by VIV of a hanging cable with a single dashpot localized at its upper end as a function of the damping parameter $\xi$ and reduced length $\ell$. The white cross locates the optimal harvesting configuration: $\xi = 1.1$ and $\ell = 68.3$, leading to an efficiency of $\eta = 0.22$.}
		\label{fig:map_string}
	\end{figure}

In this section, the VIV model is used to perform a parametrical study of the energy harvesting by VIV of a hanging cable,Figure \ref{fig:scheme} (c). The model is that given in Equations \eqref{eq:WO_corde_eff} and \eqref{eq:BC_manips}, but with $h = \infty$, corresponding to the case shown inFigure \ref{fig:scheme} (c). For the sake of comparison with the tensioned cable configuration, the model parameters are now the same as in Section \ref{subsec:Efficiency_cable}: $C_{D} = 2$, $C_{L0} = 0.8$, $St = 0.17$, $\mu = 2.79$, $A = 12$ and $\varepsilon = 0.3$. Note that the coupling parameter $A$ and the Strouhal number differ from Section \ref{subsec:hanging_manip}, where lower Reynolds numbers were considered. The efficiency map is given inFigure \ref{fig:map_string}, in terms of the parameters $\xi$ and $\ell$.

An optimal set of parameters, $\ell_{opt} = 68.3$ and $\xi_{opt} = 1.1$, is identified. The corresponding efficiency reaches $\eta_{opt} = 0.22$, which is slightly higher than for the tensioned cable, Section \ref{sec:tensioned_cable}. The dependence of $\eta$ on the harvesting parameters is much simpler than for the tensioned cable,Figure \ref{fig:Map_cable}. In particular, maximum efficiency is achieved around $\xi \approx 1$, as already noted in experiments, regardless of the value of the length parameter, $\ell$. The fir-tree shape of the efficiency map is due to the different modes that are successively excited. It is clear that in this case, the mode selection has very little impact on the efficiency, contrary to the case of a tensioned cable, Section \ref{sec:tensioned_cable}. 

\subsection{Stationary and traveling waves}

\begin{figure}[t]
		\begin{center}
%	 		\begin{minipage}{0.3\linewidth}
%			\psfrag{x5}[tc][cc][0.65]{5}
%			\psfrag{xm5}[tc][cc][0.65]{-5}
%			\psfrag{x0}[tc][cc][0.65]{0}
%			\psfrag{y0}[cc][cc][0.65]{0}
%			\psfrag{y20}[rc][cc][0.65]{20}
%			\psfrag{y40}[rc][cc][0.65]{40}
%			\psfrag{y60}[rc][cc][0.65]{60}
%			\psfrag{z}[rc][lc][0.85][-90]{$z$}
%			\psfrag{y}[tc][bc][0.85]{$y$}
%			\psfrag{a}[cc][cc][0.75]{(a)}
%			\includegraphics[width = 0.92\linewidth, height = 5cm]{Figure_10a.eps}
%			\end{minipage} 
%	 		\begin{minipage}{0.3\linewidth}
%			\psfrag{x1}[tc][cc][0.65]{1}
%			\psfrag{xm1}[tc][cc][0.65]{-1}
%			\psfrag{x0}[tc][cc][0.65]{0}
%			\psfrag{y0}[cc][cc][0.65]{0}
%			\psfrag{y20}[rc][cc][0.65]{20}
%			\psfrag{y40}[rc][cc][0.65]{40}
%			\psfrag{y60}[rc][cc][0.65]{60}
%			\psfrag{z}[rc][lc][0.85][-90]{}
%			\psfrag{y}[tc][bc][0.85]{$y$}
%			\psfrag{b}[cc][cc][0.75]{(b)}
%			\includegraphics[width = 0.92\linewidth, height = 5cm]{Figure_10b.eps}
%			\end{minipage}
%	 		\begin{minipage}{0.3\linewidth}
%			\psfrag{x5}[tc][cc][0.65]{5}
%			\psfrag{xm5}[tc][cc][0.65]{-5}
%			\psfrag{x0}[tc][cc][0.65]{0}
%			\psfrag{y0}[cc][cc][0.65]{0}
%			\psfrag{y20}[rc][cc][0.65]{20}
%			\psfrag{y40}[rc][cc][0.65]{40}
%			\psfrag{y60}[rc][cc][0.65]{60}			
%			\psfrag{z}[rc][lc][0.85][-90]{}
%			\psfrag{y}[tc][bc][0.85]{$y$}
%			\psfrag{c}[cc][cc][0.75]{(c)}
%			\includegraphics[width = 0.92\linewidth, height = 5cm]{Figure_10c.eps}
%			\end{minipage}				
\includegraphics[width=.9\textwidth]{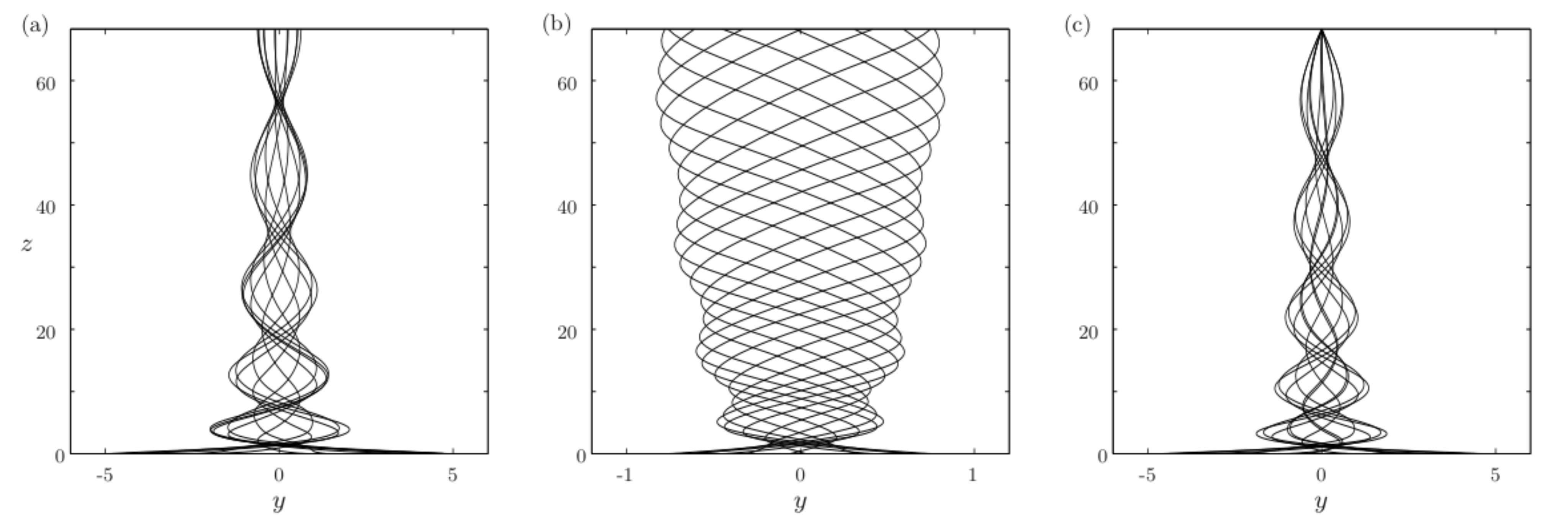}
		\end{center}		
		\caption{Selected instantaneous positions of a hanging cable VIV with a damped upper extremity for $\ell = \ell_{opt}$ and three different damping parameters : (a) $\xi = \xi_{opt} / 100$ , (b) $\xi = \xi_{opt}$ and (c) $\xi = 100 \xi_{opt}$.}
		\label{fig:string_dynamics}
	\end{figure}

In order to understand the dependence of the efficiency on $\xi$, the motion of the hanging cable is plotted inFigure \ref{fig:string_dynamics} for three different damping parameters and $\ell = \ell_{opt}$. For both small and high dampings,Figure \ref{fig:string_dynamics} (a) and (c), the motion is characterized by mainly-stationary waves, as in the case of a fixed upper extremity \citep{Grou}. The corresponding mode shape is actually very close to a zero-order Bessel function of the first kind, with either a free-end condition, $y^{\prime}\left( \ell \right) = 0$ for small $\xi$,Figure \ref{fig:string_dynamics} (a), or a fixed-end condition, $y \left( \ell \right) = 0$ for high $\xi$,Figure \ref{fig:string_dynamics} (c). However, when the optimal damping parameter is considered, $\xi = 1.1$,Figure \ref{fig:string_dynamics} (b), traveling waves appear and are oriented towards the top of the cable, where energy is extracted by the dissipative boundary condition. The local amplitude of oscillations is also more uniform than for the two other values of $\xi$, the lower end displacement being much smaller. The modification of the cable cross-flow dynamics from stationary waves to traveling waves is associated with the high efficiency reached for $\xi = 1.1$, and more generally around $\xi = 1$. In fact, the local conservation of energy for the hanging cable reads as

\begin{equation}
	\dfrac{\partial}{\partial t} \left[ \dfrac{1}{2} \dot{y}^{2} + \dfrac{1}{2} z y^{\prime}{}^{2}  \right] = M q \dot{y} - \dfrac{\gamma}{\mu} \dot{y}^{2} + \left( z y^{\prime} \dot{y} \right)^{\prime},
	\label{eq:bilan_energ}
\end{equation}

\noindent{where $M q \dot{y}$ is the energy transferred to the solid by its wake through lift, $\gamma \dot{y}^{2} / \mu$ is the energy dissipated by the flow and $\left( z y^{\prime} \dot{y} \right)^{\prime}$ is the instantaneous balance of the energy flux at location $z$. The energy flux $\phi$ along the cable consequently reads as}

 \begin{equation}
 	\phi \left( z \right) = - \left< z y^{\prime} \dot{y} \right>,
 	\label{eq:flux}
 \end{equation}
	 
\begin{figure}[t]
		\begin{center}
%	 		\begin{minipage}{0.3\linewidth}
%			\psfrag{xm2}[tc][cc][0.65]{-2}
%			\psfrag{xm1}[tc][cc][0.65]{-1}
%			\psfrag{x0}[tc][cc][0.65]{\textbf{0}}
%			\psfrag{y0}[cc][cc][0.65]{0}
%			\psfrag{y20}[rc][cc][0.65]{20}
%			\psfrag{y40}[rc][cc][0.65]{40}
%			\psfrag{y60}[rc][cc][0.65]{60}	
%			\psfrag{z}[rc][lc][0.85][-90]{$z$}
%			\psfrag{y}[tc][bc][0.85]{$\phi \left( z \right) $}
%			\psfrag{a}[cc][cc][0.75]{(a)}
%			\includegraphics[width = 0.92\linewidth, height = 5cm]{Figure_11a.eps}
%			\end{minipage} 
%	 		\begin{minipage}{0.3\linewidth}
%			\psfrag{x2}[tc][cc][0.65]{2}
%			\psfrag{x1}[tc][cc][0.65]{1}
%			\psfrag{x0}[tc][cc][0.65]{\textbf{0}}
%			\psfrag{y0}[cc][cc][0.65]{0}
%			\psfrag{y20}[rc][cc][0.65]{20}
%			\psfrag{y40}[rc][cc][0.65]{40}
%			\psfrag{y60}[rc][cc][0.65]{60}	
%			\psfrag{z}[rc][lc][0.85][-90]{}
%			\psfrag{y}[tc][bc][0.85]{$ \phi \left( z \right) $}
%			\psfrag{b}[cc][cc][0.75]{(b)}
%			\includegraphics[width = 0.92\linewidth, height = 5cm]{Figure_11b.eps}
%			\end{minipage}
%	 		\begin{minipage}{0.3\linewidth}
%			\psfrag{xm2}[tc][cc][0.65]{-2}
%			\psfrag{xm1}[tc][cc][0.65]{-1}
%			\psfrag{x0}[tc][cc][0.65]{\textbf{0}}
%			\psfrag{y0}[cc][cc][0.65]{0}
%			\psfrag{y20}[rc][cc][0.65]{20}
%			\psfrag{y40}[rc][cc][0.65]{40}
%			\psfrag{y60}[rc][cc][0.65]{60}	
%			\psfrag{z}[rc][lc][0.85][-90]{}
%			\psfrag{y}[tc][bc][0.85]{$\phi \left( z \right)$}
%			\psfrag{c}[cc][cc][0.75]{(c)}
%			\includegraphics[width = 0.92\linewidth, height = 5cm]{Figure_11c.eps}
%			\end{minipage}		
\includegraphics[width=.9\textwidth]{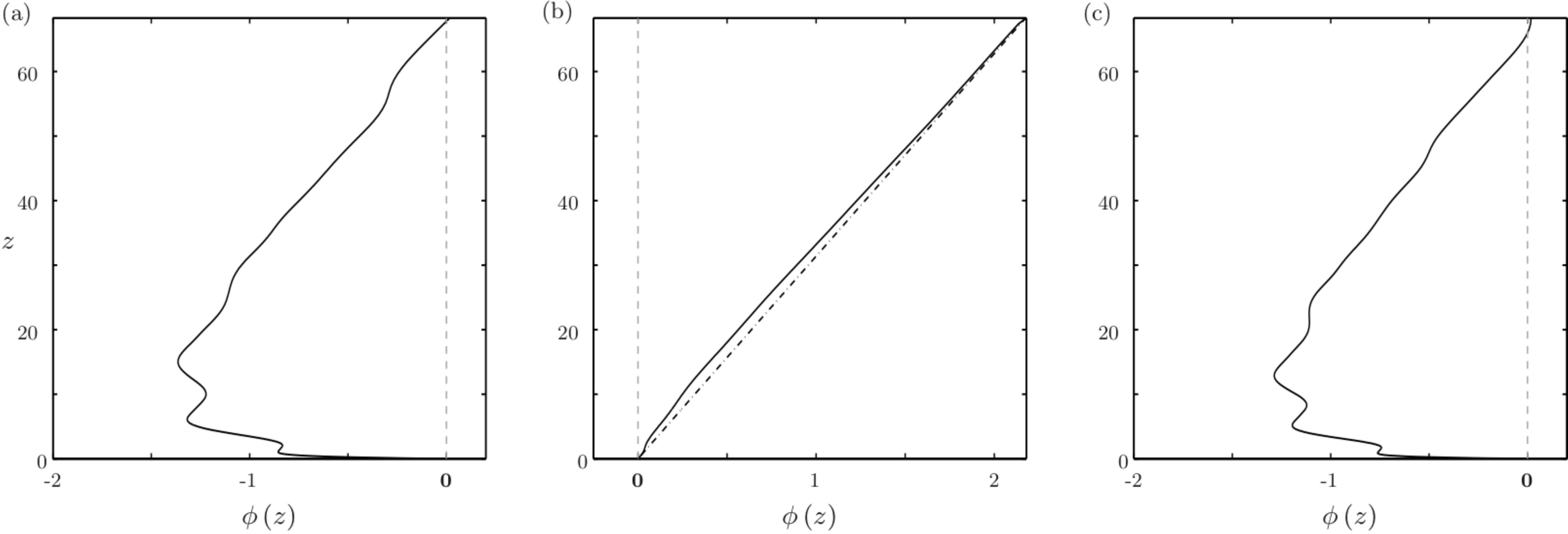}		
		\end{center}		
		\caption{Energy flux $ \phi \left( z \right) $ for (a) $\xi = \xi_{opt} / 100$ , (b) $\xi = \xi_{opt}$ and (c) $\xi = 100 \xi_{opt}$. The dashed-dotted line shows the linear approximation of $\phi$ on plot (b).}
		\label{fig:Flux}
	\end{figure}

\noindent{and its sign indicates the direction of the energy propagation along the structure. This energy flux is plotted inFigure \ref{fig:Flux}, for the three cases considered previously. For both small and high damping parameters, the flux is negative everywhere along the cable, except in the close vicinity of the dashpot,Figure \ref{fig:string_dynamics} (a) and (c); energy travels downwards and is dissipated by the flow in regions of high amplitudes of oscillations. Conversely, the flux $\phi$ is positive and increases linearly along the span of the cable for the optimal $\xi$. The energy transferred from the flow to the cable all along its span is then transported towards the energy harvester by the traveling waves. Note that for an ideal traveling wave of amplitude $y_{0}$, wavenumber $k$ and frequency $\omega$, the flux is $\phi \left( z \right) = k \omega y_{0}^{2} / 2$. This approximation of the energy flux is plotted inFigure \ref{fig:Flux} (b) and is in very good agreement with the actual energy flux. This confirms that the high efficiencies are caused by energy transport towards the damped end of the system, through traveling waves, thanks to the cable flexibility and the gravity-induced tension.}

\section{Discussion and Conclusion}	\label{sec:ccl}

Two configurations subjected to VIV have been considered in this paper, a tensioned cable with periodic harvesters (PH in the following) and a hanging cable with a top harvester (TH in the following). In both cases, predictions of efficiency were presented using a model validated against DNS and experiments. The first important result is that the optimal efficiencies of both configurations are close to that of an elastically-mounted rigid cylinder with a harvester, $\eta \approx 0.20$. These optimal efficiencies are found under lock-in, in regions of the parameter space where the vortex shedding frequency is close to the natural frequency of a structural mode. For these cable systems, the multiplicity of vibration modes allows multiple lock-ins and thereby multiple high-efficiency regions. This is a clear advantage compared to rigid systems and represents a second important result. The third main result is that these two configurations behave differently, as illustrated by the shape of the efficiency maps,Figures \ref{fig:Map_cable} and \ref{fig:map_string}. These maps were analyzed in details, and the differences shown to be related to the difference in the selection of waves by the boundary conditions. In the PH case, stationary waves dominate and the key parameter is the excited mode number, the parity of which is essential. In the TH case, traveling waves allow the transport of energy to the harvester, provided that the intensity of the harvester is properly selected.

In terms of practical applications, the two configurations behave differently by several aspects. In order to answer the question of which configuration should be installed in a flow of given velocity $U_{0}$, the optimal dimensional power generated by one harvester, $\left< \mathcal{P} \right>$, is computed using the optimal parameters identified in Sections \ref{subsec:Efficiency_cable} and \ref{subsec:eff_corde}. The mean flow velocity indeed impacts the optimal dimensional power $\left< \mathcal{P} \right>$ through the energy flux $\rho D U^{3}/2$ and the optimal value of $L$, with $L$ the distance between two harvesters for PH case and $L$ the length of the system for TH case. The optimal harvested power scales differently with the dimensional flow velocity $U_{0}$,

\begin{equation}
	\left< \mathcal{P} \right>_{\text{PH}} \sim U_{0}^{2} \quad \mbox{ and } \quad  \left< \mathcal{P} \right>_{\text{TH}} \sim U_{0}.
	\label{eq:P_scaling}
\end{equation}

\begin{figure}[!t]
	\begin{center}
%			\psfrag{xm2}[tc][cc][0.6]{$10^{-2}$}
%			\psfrag{xm1}[tc][cc][0.6]{$10^{-1}$}
%			\psfrag{x00}[tc][cc][0.6]{$10^{0}$}
%			\psfrag{x01}[tc][cc][0.6]{$10^{1}$}
%			\psfrag{ym3}[cc][lc][0.6]{$10^{-3}$}
%			\psfrag{ym2}[cc][cc][0.6]{}
%			\psfrag{ym1}[cc][lc][0.6]{$10^{-1}$}
%			\psfrag{y00}[cc][cc][0.6]{}
%			\psfrag{y01}[cc][lc][0.6]{$10^{1}$}
%			\psfrag{y02}[cc][cc][0.6]{}
%			\psfrag{y03}[cc][lc][0.6]{$10^{3}$}
%			\psfrag{y04}[cc][cc][0.6]{}
%			\psfrag{U}[tc][bc][0.85]{$U_{0}$ (m/s)}
%			\psfrag{P}[ct][lb][0.85][-90]{$\left< \mathcal{P} \right>$ (W)}
%			\psfrag{b}[cc][cc][0.65]{}
%			\psfrag{TH}[cc][cc][0.85]{TH}	
%			\psfrag{PH}[cc][cc][0.85]{PH}				
			\includegraphics[width = 0.6\linewidth]{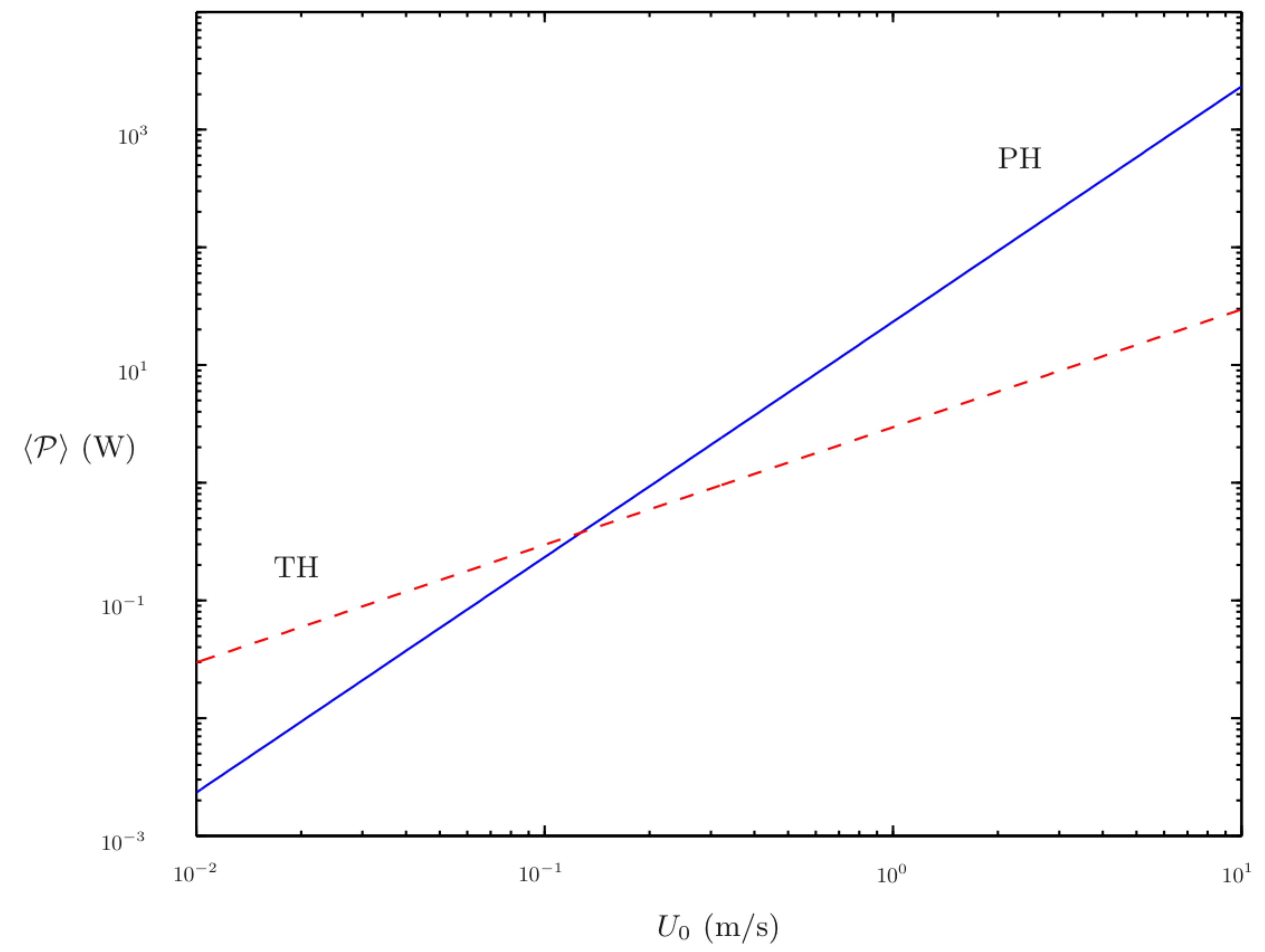}
	\end{center}
	\caption{Dimensional power, $\left< \mathcal{P} \right>$ (in W), harvested  by one harvester for VIV of a tensioned cable with periodically-distributed energy harvesters (solid line) or a hanging cable (dashed line). For each flow velocity $U_{0}$ and each configuration, the optimal dimensional parameters are derived and give access to the optimal power $\left< \mathcal{P} \right>$ which is extracted by a single harvester.}
	\label{fig:scaling}
\end{figure}

\noindent{The TH configuration seems more adapted to lower flow velocities and the PH to higher ones. This is illustrated inFigure \ref{fig:scaling}, for a given cable ($D = 0.04 \mbox{ m}$ and $m_{s} = 3.20 \mbox{ kg/m}$) and a given tension for the PH case ($\Theta = 10 \mbox{ kN}$), where the crossing point of the two configurations is near $U_{0} = 0.1 \mbox{ m/s}$. Clearly, the systems are more adapted to rather low flow velocities and thus low generated power, since the optimal power may vary like $U_{0}^{3}$ for other energy harvesting systems.}

\begin{figure}[t]	
	\begin{center}
%		\psfrag{x0}[cc][cc][0.65]{$0$}
%		\psfrag{x2}[cc][cc][0.65]{$2$}
%		\psfrag{x4}[cc][cc][0.65]{$4$}
%		\psfrag{x1}[cc][cc][0.65]{$1$}	
%		\psfrag{x3}[cc][cc][0.65]{$3$}	
%		\psfrag{x5}[cc][cc][0.65]{$5$}	
%		\psfrag{y0}[cc][cc][0.65]{$0$}			
%		\psfrag{y1}[rc][cc][0.65]{$0.1$}				
%		\psfrag{y2}[rc][cc][0.65]{$0.2$}				
%		\psfrag{U}[tc][cc][0.85]{$\dfrac{U}{U_{0}}$}	
%		\psfrag{eta}[cc][cc][0.85][-90]{$\eta$}			
		\includegraphics[width = 0.6\linewidth]{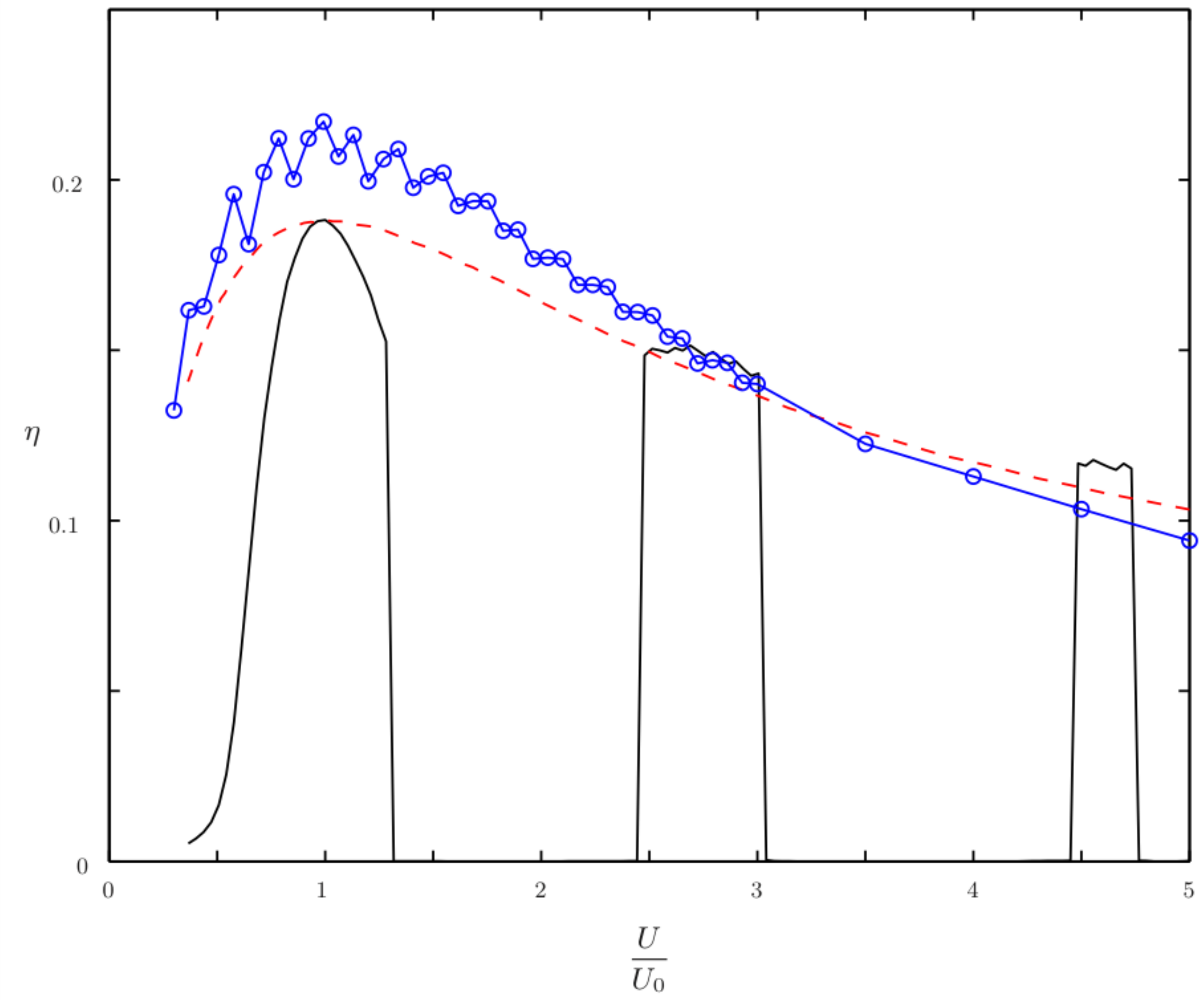}
	\end{center}
	\caption{Sensitivity of the efficiency $\eta$ to flow speed variations for the three flexible devices considered: the tensioned cable with constant tension (solid line), the tensioned cable with a drag-induced tension (dashed line) and the hanging cable (line and symbols).}
	\label{fig:sensitivity_speed}
\end{figure}

The two configurations differ in terms of their robustness versus possible variations of the flow velocity around the design point. This appears in the efficiency maps,Figures \ref{fig:Map_cable} and \ref{fig:map_string}, where the PH case shows discontinuities while the TH case exhibits smoother variations of efficiency with the parameters.Figure \ref{fig:sensitivity_speed} shows the effect on the efficiency of the actual velocity $U$, normalized by the reference value at design, $U_{0}$, for the two cases. The PH case exhibits ranges of zero efficiency, while the TH case varies more regularly. These evolutions result from the efficiency maps discussed above, knowing that $\ell \sim U$ and $\xi \sim 1/U$ in the PH case while $\ell \sim U^{2}$ and $\xi$ is independent of $U$ in the TH case. The TH case seems much more robust. Yet, a simple solution exists to regularize the efficiency of the PH case, by having a tension that is induced by a drag force, $\Theta \sim U^{2}$,Figure \ref{fig:sensitivity_speed}. In that case, the natural frequencies of the structural modes vary simultaneously with the vortex shedding frequency and lock-in is preserved on the lowest mode.

Altogether, VIV of cables with localized harvesters seem a promising way to harvest energy from low velocity geophysical flows. Considerable work, however, remains to be done to design real harvesters, but the simplicity of the flexible structures is a clear advantage in terms of building and maintenance.

%\section*{References}
%\bibliographystyle{model1-num-names}
%\bibliographystyle{elsarticle-harv}
%\bibliography{Biblio}

\begin{thebibliography}{39}
\expandafter\ifx\csname natexlab\endcsname\relax\def\natexlab#1{#1}\fi
\expandafter\ifx\csname url\endcsname\relax
  \def\url#1{\texttt{#1}}\fi
\expandafter\ifx\csname urlprefix\endcsname\relax\def\urlprefix{URL }\fi

\bibitem[{Baarholm et~al.(2006)Baarholm, Larsen, and Lie}]{Baar}
Baarholm, G., Larsen, C., Lie, H., 2006. On fatigue damage accumulation from
  in-line and cross-flow vortex-induced vibrations on risers. Journal of Fluids
  and Structures 22, 109--127.

\bibitem[{Bahaj(2013)}]{Baha}
Bahaj, A., 2013. Marine current energy conversion: the dawn of a new era in
  electricity production. Philosophical Transactions of the Royal Society A
  371, 20120500.

\bibitem[{Barrero-Gil et~al.(2010)Barrero-Gil, Alonso, and Sanz-Andres}]{Barr}
Barrero-Gil, A., Alonso, G., Sanz-Andres, A., 2010. Energy harvesting from
  transverse galloping. Journal of Sound and Vibration 329, 2873--2883.

\bibitem[{Barrero-Gil et~al.(2012)Barrero-Gil, Pindado, and Avila}]{Barr2}
Barrero-Gil, A., Pindado, S., Avila, S., 2012. Extracting energy from
  vortex-induced vibrations : a parametric study. Applied mathematical
  modelling 36, 3153--3160.

\bibitem[{Batten et~al.(2013)Batten, Harrison, and Bahaj}]{Batt}
Batten, W., Harrison, M., Bahaj, A., 2013. Accuracy of the actuator disc-{RANS}
  approach for predicting the performance and wake of tidal turbines.
  Philosophical Transactions of the Royal Society A 371, 20120293.

\bibitem[{Bernitsas et~al.(2008)Bernitsas, Raghavan, Ben-Simon, and
  Garcia}]{Bern}
Bernitsas, M.~M., Raghavan, K., Ben-Simon, Y., Garcia, E. M.~H., 2008. Vivace
  ({V}ortex-{I}nduced {V}ibration {A}quatic {C}lean {E}nergy) : A new concept
  in generation of clean and renewable energy from fluid flow. Journal of
  Offshore Mechanics and Arctic Engineering 130~(041101), 1--15.

\bibitem[{Blevins(1990)}]{Blev}
Blevins, R., 1990. Flow-induced vibration - Second edition. Van Nostrand
  Reinhold, New York.

\bibitem[{Boragno et~al.(2012)Boragno, Festa, and Mazzino}]{Bora}
Boragno, C., Festa, R., Mazzino, A., 2012. Elastically bounded flapping wing
  for energy harvesting. Applied Physics Letters 100, 253906.

\bibitem[{Bourguet et~al.(2011)Bourguet, Karniadakis, and Triantafyllou}]{Bour}
Bourguet, R., Karniadakis, G.~E., Triantafyllou, M.~S., 2011. Vortex-induced
  vibrations of a long flexible cylinder in shear flow. Journal of Fluid
  Mechanics 677, 342--382.

\bibitem[{Chaplin et~al.(2005)Chaplin, Bearman, Huera-Huarte, and
  Pattenden}]{Chap}
Chaplin, J., Bearman, P., Huera-Huarte, F., Pattenden, R., 2005. Laboratory
  measurements of vortex-induced vibrations of a vertical tension riser in a
  stepped current. Journal of Fluids and Structures 21, 3--24.

\bibitem[{de~Langre(2006)}]{Lang}
de~Langre, E., 2006. Frequency lock-in is caused by coupled-mode flutter.
  Journal of Fluids and Structures 22, 783--791.

\bibitem[{Doar\'{e} and Michelin(2011)}]{Doar}
Doar\'{e}, O., Michelin, S., 2011. Piezoelectric coupling in energy-harvesting
  fluttering flexible plates : linear stability analysis and conversion
  efficiency. Journal of Fluids and Structures 27, 1357--1375.

\bibitem[{Facchinetti et~al.(2004)Facchinetti, de~Langre, and Biolley}]{Facc}
Facchinetti, M.~L., de~Langre, E., Biolley, F., 2004. Coupling of structure and
  wake oscillators in vortex-induced vibrations. Journal of Fluids and
  Structures 19, 123--140.

\bibitem[{Grouthier(2013)}]{Grou2}
Grouthier, C., 2013. R\'{e}cup\'{e}ration d'\'{e}nergie et vibrations induites
  par vortex de structures flexibles. PhD Thesis, \'{E}cole Polytechnique.

\bibitem[{Grouthier et~al.(2013)Grouthier, Michelin, Modarres-Sadeghi, and
  de~Langre}]{Grou}
Grouthier, C., Michelin, S., Modarres-Sadeghi, Y., de~Langre, E., 2013.
  Self-similar vortex-induced vibrations of a hanging string. Journal of Fluid
  Mechanics 724, R2.

\bibitem[{Hartlen and Currie(1970)}]{Hart}
Hartlen, R., Currie, I., 1970. Lift-oscillator model for vortex-induced
  vibration. Proceddings of American Society of Civil Engineers, Journal of the
  Engineering Mechanics Divisin 96, 577--591.

\bibitem[{Hobbs and Hu(2012)}]{Hobb}
Hobbs, W., Hu, D., 2012. Tree-inspired piezoelectric energy harvesting. Journal
  of Fluids and Structures 28, 103--114.

\bibitem[{Karniadakis and Sherwin(1999)}]{Karn}
Karniadakis, G., Sherwin, S., 1999. Spectral/HP Element Methods for CFD, 1st
  edn. Oxford University Press.

\bibitem[{King(1995)}]{King}
King, R., 1995. An investigation of vortex-induced vibrations of sub-sea
  communications cables. In: Proceedings of the 6th International conference on
  Flow-Induced Vibration, London, UK : P.W. Bearman (ed). pp. 443--454.

\bibitem[{Michelin and Doar\'{e}(2013)}]{Mich}
Michelin, S., Doar\'{e}, O., 2013. Energy harvesting efficiency of
  piezoelectric flags in axial flows. Journal of Fluid Mechanics 714, 489--504.

\bibitem[{Modarres-Sadeghi et~al.(2010)Modarres-Sadeghi, Mukundan, Dahl, Hover,
  and Triantafyllou}]{Moda2}
Modarres-Sadeghi, Y., Mukundan, H., Dahl, J., Hover, F., Triantafyllou, M.,
  2010. The effect of higher harmonic forces on fatigue life of marine risers.
  Journal of Sound and Vibration 329, 43--55.

\bibitem[{Mukundan et~al.(2009)Mukundan, Modarres-Sadeghi, Dahl, Hover, and
  Triantafyllou}]{Muku}
Mukundan, H., Modarres-Sadeghi, Y., Dahl, J., Hover, F., Triantafyllou, M.,
  2009. Monitoring viv fatigue damage on marine risers. Journal of Fluids and
  Structures 25, 617--628.

\bibitem[{Naudascher and Rockwell(1990)}]{Naud}
Naudascher, E., Rockwell, D., 1990. Flow-induced vibration - an engineering
  guide. Balkema A.A.

\bibitem[{Nishino and Willden(2012)}]{Nish}
Nishino, T., Willden, R., 2012. The efficiency of an array of tidal turbines
  partially blocking a wide channel. Journal of Fluid Mechanics 708, 596--606.

\bibitem[{Norberg(2003)}]{Norb}
Norberg, C., 2003. Fluctuating lift on a circular cylinder : review and new
  measurements. Journal of Fluids and Structures 17, 57--96.

\bibitem[{Peng and Zhu(2009)}]{Peng}
Peng, Z., Zhu, Q., 2009. Energy harvesting through flow-induced oscillations of
  a foil. Physics of Fluids 21, 123602.

\bibitem[{Singh et~al.(2012{\natexlab{a}})Singh, Michelin, and
  de~Langre}]{Sing2}
Singh, K., Michelin, S., de~Langre, E., 2012{\natexlab{a}}. The effect of
  non-uniform damping on flutter in axial flow and energy-harvesting
  strategies. Proceedings of the Royal Society A 468, 3620--3635.

\bibitem[{Singh et~al.(2012{\natexlab{b}})Singh, Michelin, and
  de~Langre}]{Sing}
Singh, K., Michelin, S., de~Langre, E., 2012{\natexlab{b}}. Energy harvesting
  from axial fluid-elastic instabilities of a cylinder. Journal of Fluids and
  Structures 30, 159--172.

\bibitem[{Skop and Balasubramanian(1997)}]{Skop2}
Skop, R., Balasubramanian, S., 1997. A new twist on an old model for
  vortex-excited vibrations. Journal of Fluids and Structures 11, 395--412.

\bibitem[{Srinil(2010)}]{Srin}
Srinil, N., 2010. Multi-mode interactions in vortex-induced vibrations of
  flexible curved/straight structures with geometric nonlinearities. Journal of
  Fluids and Structures 26, 1098--1122.

\bibitem[{Tang et~al.(2009)Tang, Pa\"{i}doussis, and Jiang}]{Tang}
Tang, L., Pa\"{i}doussis, M.~P., Jiang, J., 2009. Cantilevered flexible plates
  in axial flow : Energy transfer and the concept of flutter mill. Journal of
  Sound and Vibration 326, 263--276.

\bibitem[{Tognarelli et~al.(2008)Tognarelli, Taggart, and Campbell}]{Togn}
Tognarelli, M., Taggart, S., Campbell, M., 2008. Actual viv response of full
  scale drilling risers: with and without supression devices. In: Proceedings
  of the ASME 27th International Conference on Offshore Mechanics and Arctic
  Engineering, Estoril, Portugal. pp. OMAE2008--57046.

\bibitem[{Vandiver(1993)}]{Vand2}
Vandiver, J., 1993. Dimensionless parameters important to the prediction of
  vortex-induced vibration of long, flexible cylinders in ocean currents.
  Journal of Fluids and Structures 7, 423--455.

\bibitem[{Violette et~al.(2007)Violette, de~Langre, and Szydlowski}]{Viol}
Violette, R., de~Langre, E., Szydlowski, J., 2007. Computations of
  vortex-induced vibrations of long structures using a wake oscilator model :
  Comparison with {DNS} and experiments. Computers and Structures 85,
  1134--1141.

\bibitem[{Violette et~al.(2010)Violette, de~Langre, and Szydlowski}]{Viol2}
Violette, R., de~Langre, E., Szydlowski, J., 2010. A linear stability approach
  to vortex-induced vibrations and waves. Journal of Fluids and Structures
  26~(3), 442--466.

\bibitem[{Williamson and Govardhan(2004)}]{Will}
Williamson, C., Govardhan, R., 2004. Vortex-induced vibrations. {Annual Review
  of Fluid Mechanics} 36, 413--455.

\bibitem[{Xu et~al.(2008)Xu, Zeng, and Wu}]{Xu}
Xu, W.-H., Zeng, X.-H., Wu, Y.-X., 2008. High aspect ratio ({L/D}) riser {VIV}
  prediction using wake oscillator model. Ocean Engineering 35, 1769--1774.

\bibitem[{Yoshitake et~al.(2004)Yoshitake, Sueoka, Yamasaki, Sugimura, and
  Ohishi}]{Yosh}
Yoshitake, Y., Sueoka, A., Yamasaki, M., Sugimura, Y., Ohishi, T., 2004.
  Quenching of vortex-induced vibrations of towering structure and generation
  of electricity using hula-hoops. Journal of Sound and Vibration 272, 21--38.

\bibitem[{Zhu et~al.(2009)Zhu, Haase, and h.~Wu}]{Zhu}
Zhu, Q., Haase, M., h.~Wu, C., 2009. Modeling the capacity of a novel
  flow-energy harvester. Applied Mathematical Modelling 33, 2207--2217.

\end{thebibliography}

\end{document}